\documentclass[a4paper,fleqn,usenatbib]{mnras}

\usepackage{Times}

\usepackage[T1]{fontenc}
\usepackage{ae,aecompl}

\usepackage{graphicx}	
\usepackage{amsmath}	
\usepackage{amssymb}	

\title[The Ly$\alpha$ escape fraction of SFGs at $z\sim2$]{The CALYMHA survey: Ly$\alpha$ escape fraction and its dependence on galaxy properties at $\bf z=2.23$}

\author[J. Matthee et al.]{Jorryt Matthee$^{1}$\thanks{E-mail: matthee@strw.leidenuniv.nl}, David Sobral$^{1,2,3}$, Iv\'an Oteo$^{4,5}$, Philip Best$^{4}$, Ian Smail$^{6}$, \newauthor Huub R\"ottgering$^{1}$ and Ana Paulino-Afonso$^{2}$ \\
$^{1}$ Leiden Observatory, Leiden University, P.O.\ Box 9513, NL-2300 RA Leiden, The Netherlands\\
$^{2}$ Instituto de Astrof\'{\i}sica e Ci\^{e}ncias do Espa\c{c}o, Universidade de Lisboa, OAL, Tapada da Ajuda, PT1349-018 Lisboa, Portugal \\
$^{3}$ Department of Physics, Lancaster University, Lancaster, LA1 4YB, UK \\ 
$^{4}$ Institute for Astronomy, University of Edinburgh, Royal Observatory, Blackford Hill, Edinburgh EH9 3HJ UK\\
$^{5}$ European Southern Observatory, Karl-Schwarzschild-Str. 2, 85748 Garching, Germany \\
$^{6}$ Centre for Extragalactic Astronomy, Department of Physics, Durham University, South Road, Durham, DH1 3LE, UK\\}

\date{Accepted 2016 February 08. Received 2016 February 08; in original form 2015 December 23}

\pubyear{2016}

\begin{document}
\label{firstpage}
\pagerange{\pageref{firstpage}--\pageref{lastpage}}
\maketitle

\begin{abstract}
We present the first results from our CAlibrating LYMan-$\alpha$ with H$\alpha$ (CALYMHA) pilot survey at the Isaac Newton Telescope. We measure Ly$\alpha$ emission for 488 H$\alpha$ selected galaxies at $z=2.23$ from HiZELS in the COSMOS and UDS fields with a specially designed narrow-band filter ($\lambda_c$ = 3918 {\AA}, $\Delta\lambda$= 52 {\AA}). We find 17 dual H$\alpha$-Ly$\alpha$ emitters ($f_{\rm Ly\alpha} >5\times10^{-17}$ erg s$^{-1}$ cm$^{-2}$, of which 5 are X-ray AGN). For star-forming galaxies, we find a range of Ly$\alpha$ escape fractions (f$_{\rm esc}$, measured with 3$''$ apertures) from $2$\%$-30$\%. These galaxies have masses from $3\times10^8$ M$_{\odot}$ to 10$^{11}$ M$_{\odot}$ and dust attenuations E$(B-V)=0-0.5$. Using stacking, we measure a median escape fraction of $1.6\pm0.5$\% ($4.0\pm1.0$\% without correcting H$\alpha$ for dust), but show that this depends on galaxy properties. The stacked f$_{\rm esc}$ tends to decrease with increasing SFR and dust attenuation. However, at the highest masses and dust attenuations, we detect individual galaxies with f$_{\rm esc}$ much higher than the typical values from stacking, indicating significant scatter in the values of f$_{\rm esc}$. Relations between f$_{\rm esc}$ and UV slope are bimodal, with high f$_{\rm esc}$ for either the bluest or reddest galaxies. We speculate that this bimodality and large scatter in the values of f$_{\rm esc}$ is due to additional physical mechanisms such as outflows facilitating f$_{\rm esc}$ for dusty/massive systems. Ly$\alpha$ is significantly more extended than H$\alpha$ and the UV. f$_{\rm esc}$ continues to increase up to at least 20 kpc (3$\sigma$, 40 kpc [2$\sigma$]) for typical SFGs and thus the aperture is the most important predictor of f$_{\rm esc}$.
\end{abstract}

\begin{keywords}
galaxies: high-redshift -- galaxies: evolution -- galaxies: ISM \end{keywords}



\section{Introduction}
The Lyman-$\alpha$ (Ly$\alpha$) emission line (rest-frame 1216 {\AA}) has emerged as a powerful tool to study distant galaxies, since it is intrinsically the brightest emission line in {\sc Hii} regions and redshifted into optical wavelengths at $z>2$. As a result, the Ly$\alpha$ line has been used to spectroscopically confirm the highest redshift galaxies \citep{Oesch2015,Zitrin2015}, select samples of galaxies with narrow-band filters \citep[e.g.][]{Ouchi2008,Matthee2015}, find extremely young galaxies \citep[e.g.][]{Kashikawa2012,Sobral2015CR7}, study the interstellar, circumgalactic and intergalactic medium \citep[e.g.][]{Rottgering1995,Cantalupo2014,Swinbank2015} and probe the epoch of reionization \citep[e.g.][]{Ouchi2010,DijkstraReview}.

However, due to the resonant nature of Ly$\alpha$, it is unknown what the observed strength of the Ly$\alpha$ emission line actually traces. While Ly$\alpha$ photons are emitted as recombination radiation in {\sc Hii} regions, where ionising photons originate from star-formation or AGN activity, Ly$\alpha$ photons can also be emitted by collisional ionisation due to cooling \citep[e.g.][]{Rosdahl2012} and shocks. Most importantly, only a small amount of neutral hydrogen is needed to get an optical depth of 1 (with column densities of $\sim 10^{14}$ cm$^{-2}$; \citealt{HayesReview}). Therefore, Ly$\alpha$ photons are likely to undergo numerous scattering events before escaping a galaxy. This increases the likelihood of Ly$\alpha$ being absorbed by dust and also leads to a lower surface brightness (detectable as Ly$\alpha$ haloes, e.g. \citealt{Steidel2011,Momose2014,Wisotzki2015}) and diffusion in wavelength space, altering line profiles \citep[e.g.][]{Verhamme2008,DijkstraReview}. 

In order to use Ly$\alpha$ to search for and study galaxies in the early Universe, it is of key importance to directly measure the fraction of intrinsically produced Ly$\alpha$ (the Ly$\alpha$ escape fraction, f$_{\rm esc}$), and to understand how that may depend on galaxy properties. Under the assumption of case B recombination radiation, f$_{\rm esc}$ can be measured by comparing the Ly$\alpha$ flux with H$\alpha$. H$\alpha$ (rest-frame 6563 {\AA}) is not a resonant line and typically only mildly affected by dust, in a well understood way \citep[e.g.][]{GarnBest2010}. Measurements of both Ly$\alpha$ and H$\alpha$ can thus improve the understanding of what Ly$\alpha$ actually traces by comparing f$_{\rm esc}$ with other observables as mass, dust content, kinematics, or Ly$\alpha$ line properties (such as the Equivalent Width (EW) and profile). 

It is in principle possible to estimate the intrinsic Ly$\alpha$ production using other tracers of the ionising photon production rate (i.e. other star formation rate (SFR) indicators). However, these all come with their own uncertainties and assumptions. For example, studies using H$\beta$ \citep[e.g.][]{Ciardullo2014} and UV selected samples \citep[e.g.][]{Gronwall2007,Nilsson2009,Blanc2011,Cassata2015} suffer from more significant and uncertain dust corrections, and may select a population which tends to be less dusty \citep[e.g.][]{Oteo2015}. UV based studies are furthermore dependent on uncertainties regarding SED modelling, and on assumptions on the time-scales (UV typically traces SFR activity over a 10 times longer timescale than nebular emission lines, e.g. \citealt{Boquien2014}). Estimates using the far-infrared \citep{Wardlow2014,Kusakabe2014} suffer from even larger assumptions on the time-scales. Remarkably though, most studies find a consistent value of f$_{\rm esc} \sim 30$\% for Ly$\alpha$ emitters (LAEs), and lower for UV selected galaxies, $\sim3-5$\% \citep[e.g.][]{Hayes2011}.

Locally, it has been found that the Ly$\alpha$ escape fraction anti-correlates with dust attenuation \citep{Cowie2010,Atek2014}, although the large scatter indicates that there are other regulators of Ly$\alpha$ escape, such as outflows \citep[e.g.][]{Kunth1998,Atek2008,RiveraThorsen2015}. However, these locally studied galaxies have been selected in different ways than typical high redshift galaxies. Green pea galaxies (selected by their strong nebular {\sc[Oiii]} emission) have recently been studied as local analogs for high redshift LAEs \citep[e.g.][]{Henry2015,Yang2015}. These studies find indications that the escape fraction correlates with HI column density, and that is also related to galactic outflows and dust attenuation. However, the sample sizes and the dynamic range are still significantly limited.

At higher redshift, it is challenging to measure the Ly$\alpha$ escape fraction, as H$\alpha$ can only be observed up to $z\sim2.8$ from the ground, while Ly$\alpha$ is hard to observe at $z<2$. Therefore, $z\sim2.5$ is basically the only redshift window where we can directly measure both Ly$\alpha$ and H$\alpha$ with current instrumentation. This experiment has been performed by \cite{Hayes2010}, who found a global average escape fraction of $5\pm4$ \%. The escape fraction is obtained by comparing integrated H$\alpha$ and Ly$\alpha$ luminosity functions (see also \citealt{Hayes2011}), so the results depend on assumptions on the shape of the luminosity function, integration limits, etc. Recently, \cite{Oteo2015} found that only 4.5\% of the H$\alpha$ emitters covered by \cite{Nilsson2009} are detected as LAEs, indicating a similar escape fraction.

\begin{figure}
	\includegraphics[width=\columnwidth]{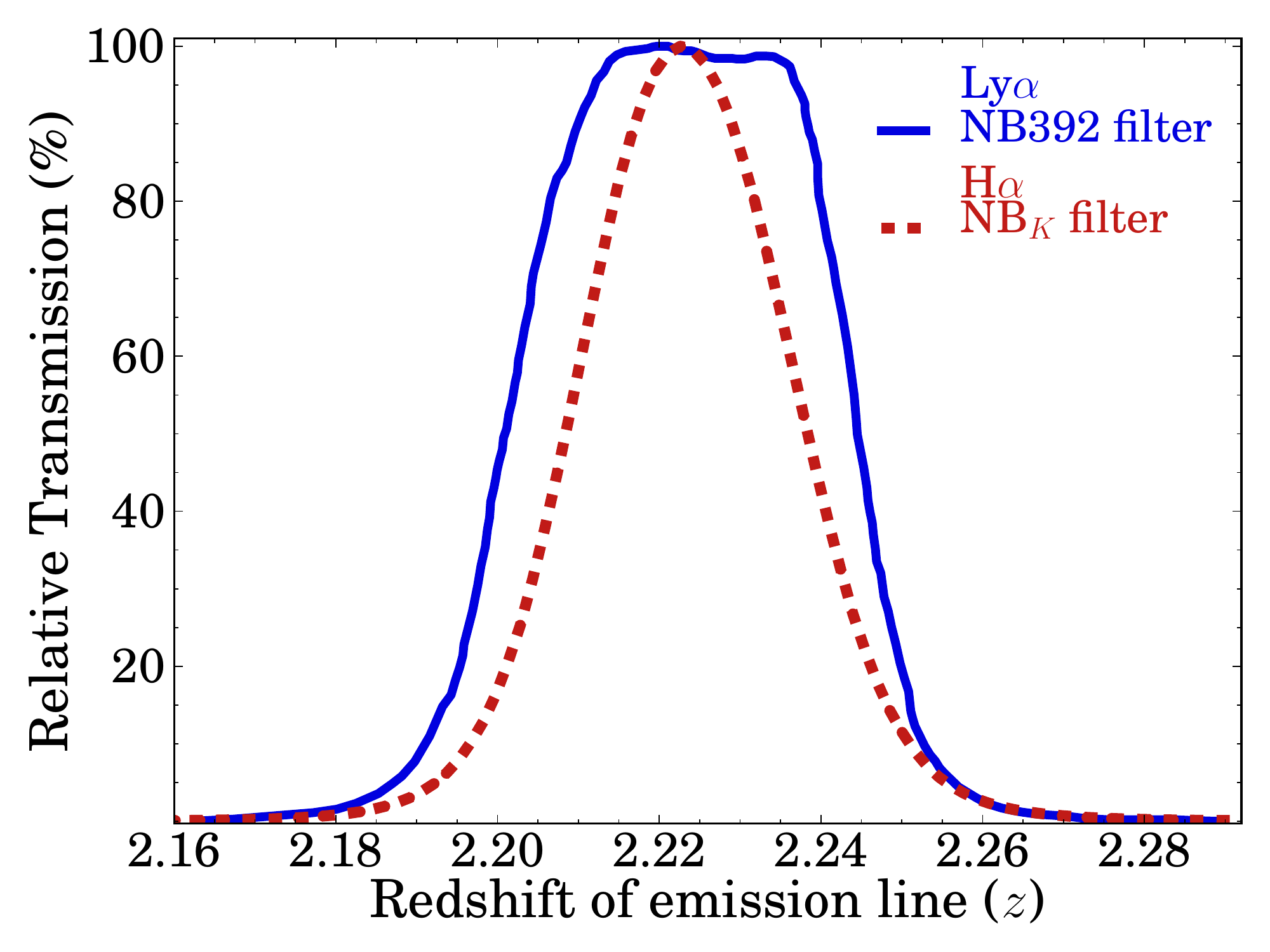}
    \caption{Filter transmission curves for the NBs used to measure H$\alpha$ (NB$_K$) and Ly$\alpha$ (NB392). The NB392 filter is designed to provide complete coverage of the redshifts at which H$\alpha$ emitters can be selected in NB$_K$, from $z=2.20-2.25$. The Ly$\alpha$ emission for all our HAEs is covered even if it is shifted by $\pm 600$ km s$^{-1}$. Depending on the specific redshift, the filter transmission varies between the two lines, such that Ly$\alpha$ is typically over-estimated with respect to H$\alpha$, see \S 3.4.2. We statistically correct for this in stacked or median measurements.}
    \label{fig:Filters}
\end{figure}

In order to increase the sample size and study dependencies on galaxy properties, we have recently completed the first phase of our CALYMHA survey: CAlibrating LYMan-$\alpha$ with H$\alpha$. This survey combines the $z=2.23$ H$\alpha$ emitters from HiZELS \citep{Sobral2013} with Ly$\alpha$ measurements using a custom-made NB filter (see Fig. $\ref{fig:Filters}$). The observations from our pilot survey presented here cover the full COSMOS field and a major part of the UDS field, and are described in Sobral et al. (in prep.).
The aim of this paper is to measure the escape fraction for the H$\alpha$ selected sources, and measure median stacked escape fractions in multiple subsets in order to understand which galaxy properties influence f$_{\rm esc}$. 

The structure of this paper is as follows. In \S 2 we present the sample of $z=2.23$ H$\alpha$ emitters and the Ly$\alpha$ observations. We describe our method to measure Ly$\alpha$ line-flux and escape fraction and galaxy properties in \S 3, while \S 4 describes our stacking method. \S 5 presents the Ly$\alpha$ properties of individual galaxies. We explore correlations between f$_{\rm esc}$ and galaxy properties in \S 6 and study extended Ly$\alpha$ emission in \S 7. Our results are compared with other studies in \S 8 and we summarise our results and present our conclusions in \S 9. 
\begin{figure*}
	\includegraphics[width=16cm]{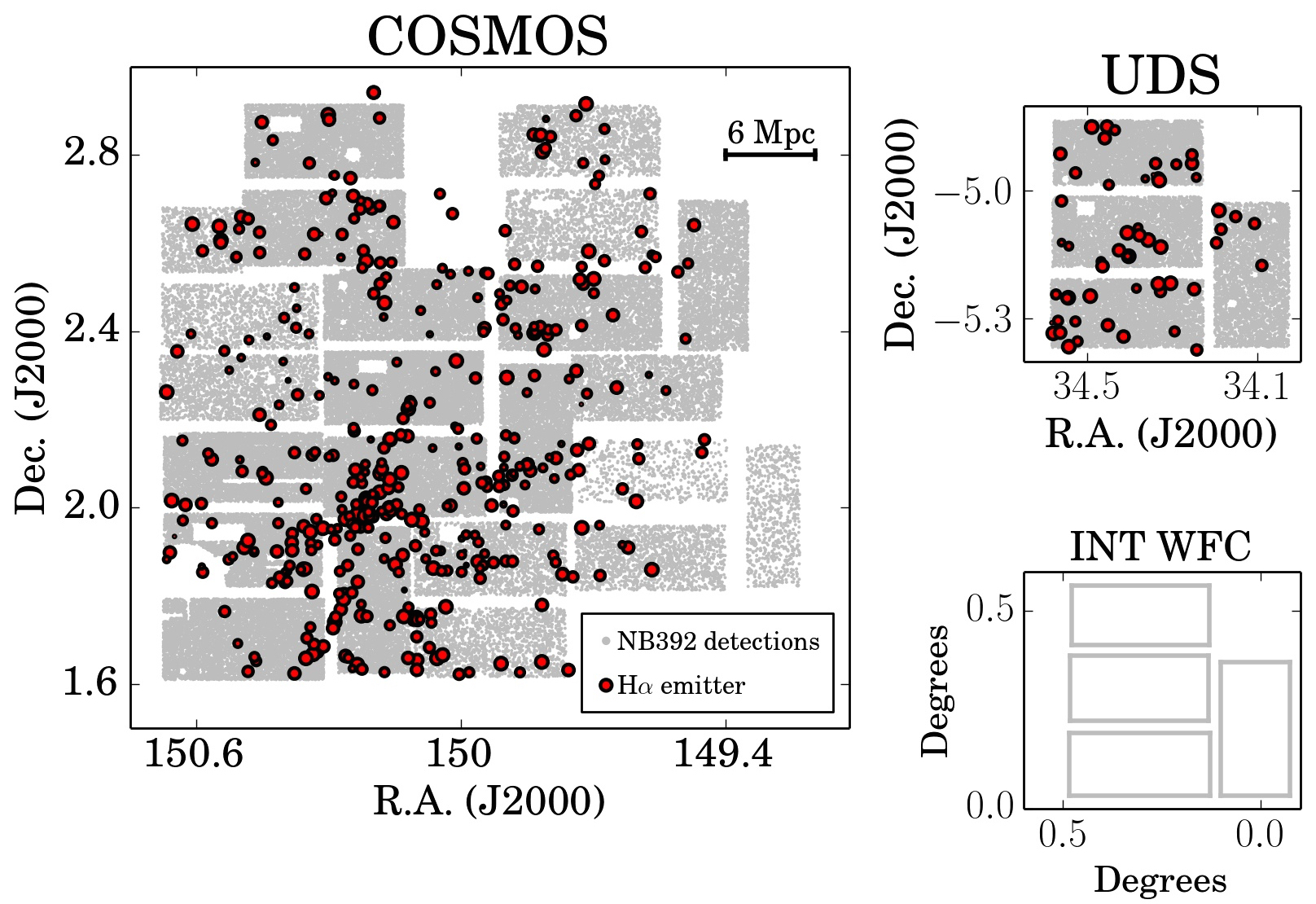}
    \caption{Positions on the sky in COSMOS and UDS of the H$\alpha$ emitters from \citet{Sobral2013} in red points, where the size of the symbols scales with observed H$\alpha$ luminosity. Our $\sim 2$ deg$^2$ coverage includes a wide range of environments, with number density of sources on the sky varying over orders of magnitudes, overcoming cosmic variance (see e.g. \citealt{Sobral2015}). The grey points show all detections in our NB392 observations, after conservative masking of noisy regions due to the dithering pattern. It can be seen that some pointings are shallower with a lower number density of sources, and that we masked regions around bright stars and severe damages to one of the chips. After our conservative masking, we use a total area of 1.208 deg$^2$ in COSMOS and 0.224 deg$^2$ in UDS. We also show the four detector chips of the WFC on the INT with a total field of view of $\sim0.25$ deg$^2$.}
    \label{fig:RADEC}
\end{figure*}

Throughout the paper, we use a $\Lambda$CDM cosmology with $H_0$ = 70 km s$^{-1} $Mpc$^{-1}$, $\Omega_{\rm M} = 0.3$ and $\Omega_{\Lambda} = 0.7$. Magnitudes are given in the AB system and measured in 3$''$ diameter apertures, unless noted otherwise. At $z=2.23$, 1$''$ corresponds to a physical scale of 8.2 kpc. We use a \cite{Chabrier2003} IMF to obtain stellar masses and star formation rates.

\section{Sample and Observations}
\subsection{Sample of H$\alpha$ emitters}
We use a sample of H$\alpha$ emitters (HAEs) at $z=2.23$ in the COSMOS and UDS fields selected from the High-$z$ Emission Line Survey (HiZELS; \citealt{Geach2008,Sobral2009,Best2013,Sobral2013}) using narrow-band (NB) imaging in the $K$ band with UKIRT. HAEs are identified using $BzK$ and $BRU$ colours and photometric redshifts, as described in \cite{Sobral2013}. These HAEs are selected to have EW$_{0, \rm H\alpha+[NII]} > 25$ {\AA}. In total, there are 588 H$\alpha$ emitters at $z=2.23$ in COSMOS, of which 552 are covered by our Ly$\alpha$ survey area. We remove 119 HAEs because they are found in noisy regions of the Ly$\alpha$ coverage, resulting in a sample of 433 HAEs in COSMOS. 
The UDS sample consists of 184 HAEs, of which 55 are observed to sufficient S/N in the INT imaging (local background of 23.5, 3$\sigma$, or deeper).
This means that our total sample includes 488 HAEs, shown in Fig. $\ref{fig:RADEC}$. 

The multi-wavelength properties of the HAEs are discussed in \cite{Oteo2015}, showing that the H$\alpha$ selection incorporates the full diversity of star-forming galaxies (e.g. in their Fig. 5 and 6), while selections based on the Lyman break or the Ly$\alpha$ emission line miss significant parts of the star-forming galaxy population at $z=2.23$. Furthermore, although our sample of galaxies contains strongly star-bursting systems, the majority is not biased towards these rare sources. Our sample is dominated by typical galaxies which are on the main relation between stellar mass and SFR (see Fig. 10 in \citealt{Oteo2015}, and e.g. \citealt{Rodigiero2014}).

\subsection{Ly$\alpha$ observations at $z=2.23$}
Ly$\alpha$ observations were conducted at the Isaac Newton Telescope (INT) at the Observatorio Roque de los Muchachos on the island of La Palma with a specially designed NB filter for the Wide Field Camera (WFC). This NB filter (NB392, $\lambda_c = 3918${\AA}, $\Delta\lambda = 52${\AA}) was designed for our survey such that it observes Ly$\alpha$ emission for all redshifts\footnote{Note that we investigate the effect of different filter transmissions between Ly$\alpha$ and H$\alpha$ as a function of redshift and the effect of systematic velocity offsets between the lines in \S 3.4.2.} at which H$\alpha$ emitters can be selected with the NB$_K$ filter, see Fig. $\ref{fig:Filters}$.

The details of the observations, data reduction and calibration are presented in Sobral et al. (2016, in prep.), where we also present the Ly$\alpha$ luminosity function (LF), and other line-emitters detected in our NB data, such as {\sc Civ$_{1549}$} at $z\approx1.5$ and {\sc[Oii]} at $z\approx0.05$. For the purpose of this paper, we use the INT observations to measure the Ly$\alpha$ flux from H$\alpha$ selected galaxies, by creating thumbnail images in NB392. For continuum estimation in COSMOS, we align publicly available $U$ and $B$ bands (from CFHT and Subaru respectively, \citealt{Capak2007,McCracken2010}) and measure the flux in these filters at the positions at which the H$\alpha$ emission is detected. In UDS, we use CFHT $U$ band data (PI: Almaini \& Foucaud) from UKIDSS UDS \citep{Lawrence2007} and Subaru $B$ band data from SXDS \citep{Furusawa2008}.

We converted the $U$, $B$, NB$_K$ and $K$ images to the pixel scale of the INT WFC (0.33$''$/pixel). The astrometry of COSMOS images is aligned using {\sc Scamp} \citep{SCAMP}, with a reference coordinate system based on {\it HST} ACS F814W band observations (as in the public COSMOS data, \citealt{McCracken2010}). The UDS images are aligned to 2MASS \citep{Skrutskie2006}. The accuracy of the astrometry is of the order of 0.1$''$. We match the full width half maximum (FWHM) of the point spread function (PSF) of all images to the FWHM of the NB392 observations (ranging from $1.8-2.0''$, depending on the particular pointing). The FWHM of reference stars was measured with {\sc SExtractor} \citep{Bertin1996}, which fits a gaussian profile to the upper 80\% of the light profile from each detected object. For NB392 imaging, we selected reference stars with magnitudes ranging from 16-18, resulting in $\sim20$ stars per WFC detector. The reference stars in $U(B)$ are fainter because in $U(B)$ stars with magnitude $<18(19)$ are saturated. For each frame, we find $\sim 50$ reference stars with magnitudes ranging from 19-21. PSF matching was then done by convolving images with a gaussian kernel. This procedure is based on the PSF matching procedure from the Subaru Suprime-Cam data reduction pipeline \citep{Ouchi2004RED}.

\section{Measurements}
\subsection{Choice of aperture}
Due to resonant scattering of Ly$\alpha$ photons, the choice of aperture can have an important consequence on the measured Ly$\alpha$ flux and escape fraction, particularly given the evidence of extended Ly$\alpha$ emission for a range of star-forming galaxies (e.g. \citealt{Steidel2011,Momose2014} and which we confirm for our sample in \S 7). Previous surveys of Ly$\alpha$ emitters typically used {\sc mag-auto} photometry with {\sc SExtractor} to measure Ly$\alpha$ fluxes \citep[e.g.][]{Hayes2010,Ouchi2010}. However, the measured flux with {\sc mag-auto} will be dependent on the depth of the NB imaging. As we are measuring Ly$\alpha$ emission for H$\alpha$ selected galaxies at the position of H$\alpha$ detection, it is impossible to perform a similar {\sc mag-auto} measurement as Ly$\alpha$ selected surveys without uncontrolled bias. This is because we have no a priori knowledge of the optimal aperture to measure Ly$\alpha$. In fact, we find in \S 5 that most H$\alpha$ emitters are undetected in Ly$\alpha$ at the flux limit of our observations. We also note that {\sc mag-auto} measurements are dependent on the depth, and therefore are not suitable for an optimal comparison as the depth of our survey varies across the field and is different than other surveys.

Due to these considerations, we choose to use a fixed diameter aperture measurements for individual sources. An aperture size of 3$''$ was chosen for the following reasons. First, it corresponds to a radial distance of 12 kpc, which is larger than the exponential scale length of Ly$\alpha$ selected sources at $z=2.2-6.6$ of $5-10$ kpc \citep{Momose2014}, and which is also similar to the reference scale used in the study of individual Ly$\alpha$ haloes (\citealt{Wisotzki2015}; although note that this survey has detected extended Ly$\alpha$ emission up to a radial distance of 25 kpc). Secondly, we find that 3$''$ aperture magnitudes on the PSF convolved images of the $U$, $B$, NB$_K$ and $K$ band recover similar magnitudes as the 2$''$ diameter apertures on the original H$\alpha$ images (which typically have a PSF FWHM $\sim 0.8''$), with a standard deviation of 0.2 magnitudes. These magnitudes from 2$''$ aperture measurements are used on most studies of the H$\alpha$ emitters from our sample \citep[e.g.][]{Sobral2013,Oteo2015}. For stacks of subsets of HAEs we vary the aperture, and discuss the difference in \S 7.

\subsection{Measuring line-fluxes}
We use fluxes in NB392, $U$ and $B$ band to measure the Ly$\alpha$ line-flux on the positions of the H$\alpha$ emitters using dual-mode {\sc SExtractor}. 
The NB392 flux is calibrated on $U$ band magnitudes of photometrically selected galaxies (see Sobral et al. 2016, in prep.), since stars have the strong Ca{\sc ii}$_{3933}$ absorption feature at the wavelengths of the NB filter. After this calibration, we also make sure that the NB excess ($U-{\rm NB}392$) is not a function of the $U-B$ colour, such that a very blue/red continuum does not bias line-flux measurements. This means that we empirically correct the NB magnitude using:
\begin{equation}
{\rm NB}392_{\rm corrected} = {\rm NB}392 + 0.19\times(U-B)-0.09.
\end{equation}
This correction ensures that a zero NB excess translates into a zero line-flux in NB392. For sources which are undetected in $U$ or $B$ we assign the median correction of the sources which are detected in $U$ and $B$, which is $+0.02$. In the following, we refer to the broadband $U$ as BB.
Then, with the NB and continuum measurements, the Ly$\alpha$ line-flux is calculated using:
\begin{equation}
f_{\rm Ly\alpha}= \Delta\lambda_{NB} \frac{f_{NB}-f_{BB}}{1-\frac{\Delta\lambda_{NB}}{\Delta\lambda_{BB}}}.
\end{equation}
Here, $f_{NB}$ and $f_{BB}$ are the flux-densities in NB392 and $U$ and $\Delta\lambda_{NB}$ and $\Delta\lambda_{BB}$ the filter-widths, which are 52 {\AA} and 758 {\AA} respectively. 

\begin{figure}
	\includegraphics[width=\columnwidth]{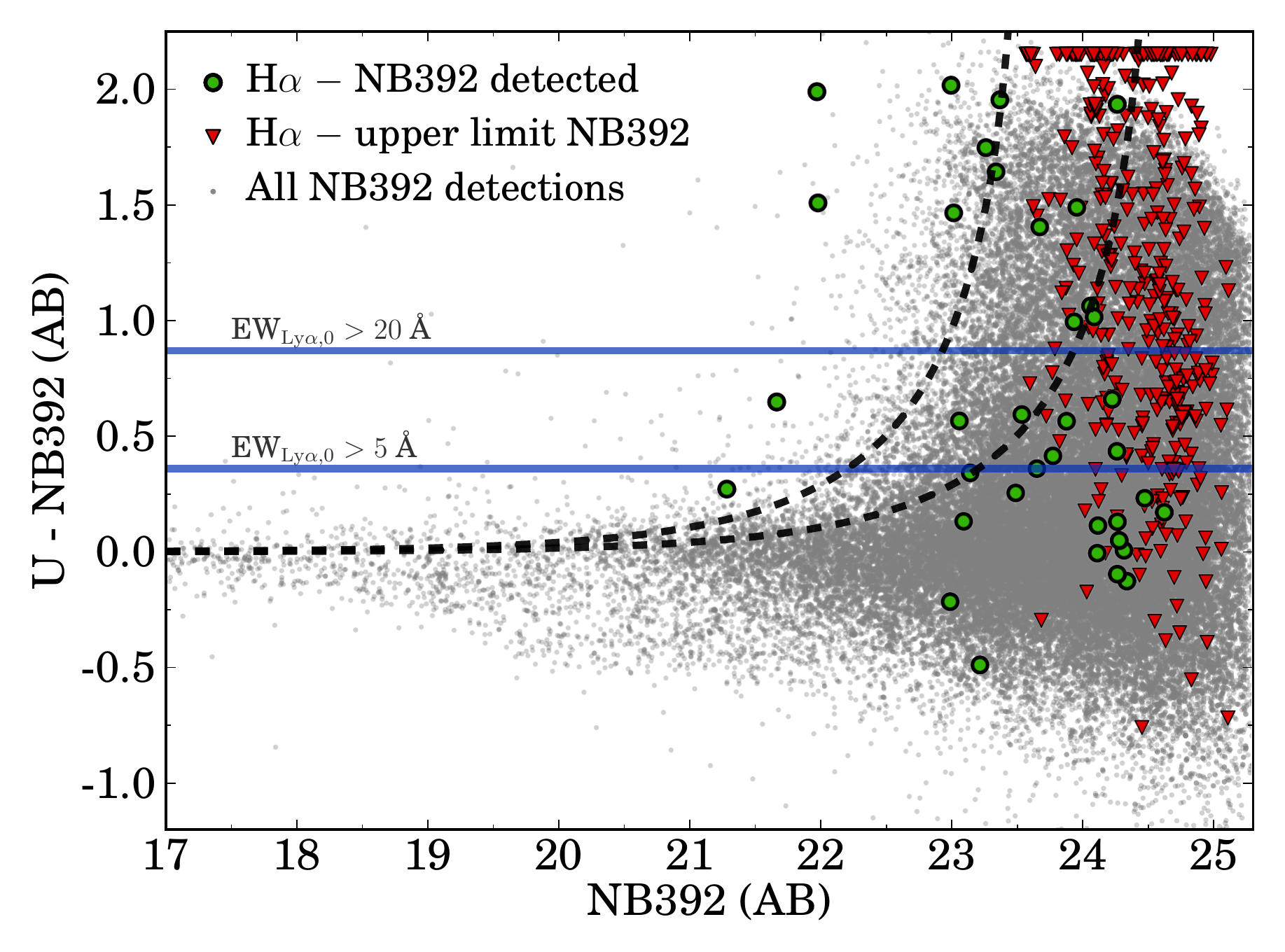}
    \caption{NB excess diagram of the sources in COSMOS and UDS. Grey points show all NB392 detections, where U has been measured in dual-mode. The green points show the H$\alpha$ emitters which are directly detected in the NB392 imaging, with measurements done at the position of the HAEs. The red triangles are upper limits at the positions of the HAEs. The blue horizontal lines show to which rest-frame Ly$\alpha$ EW a certain excess corresponds. Dashed black lines show the excess significance for either the shallowest (left) or deepest (right) NB392 data. Note that some upper limits on the NB392 magnitude are actually weaker than some detections. This is due to variations in the depth of our NB392 observations across the field. Many stars have a negative excess due to the Ca{\sc ii}$_{3933}$ absorption feature.}
    \label{fig:excess}
\end{figure}

We measure H$\alpha$ line-fluxes as described in \cite{Sobral2013}. The relevant NB is NB$_K$ and the continuum is measured in $K$ band. The excess is corrected with the median correction of $+0.03$ derived from $H-K$ colours. For an HAE to be selected as a double H$\alpha$-Ly$\alpha$ emitter, we require the $U-{\rm NB}392$ excess to be $>0.2$ (corresponding to EW$_0 > 4$ {\AA}) and a Ly$\alpha$ excess significance $\Sigma > 2$ \citep[c.f.][]{Bunker1995,Sobral2013}, see the dashed lines in Fig. $\ref{fig:excess}$, which we base on local measurements of the NB and broadband background in empty 3$''$ diameter apertures. This relatively low excess significance is only appropriate because we observe pre-selected H$\alpha$ emitters. We note however that all our directly detected sources are detected with at least 3$\sigma$ significance in the NB392 imaging.

\subsection{Measuring the Ly$\alpha$ escape fraction}
In order to measure the observed fraction of Ly$\alpha$ flux, we need to carefully estimate the intrinsic Ly$\alpha$ line-flux. The intrinsic emission of Ly$\alpha$ due to recombination radiation is related to the H$\alpha$ flux, and scales with the number of ionising photons per second. Assuming case B recombination, a temperature $T$ between 5,000 $-$ 20,000K and electron density $n_e$ ranging from $10^2 - 10^4$cm$^{-3}$, the intrinsic ratio of Ly$\alpha$/H$\alpha$ ranges from 8.1-11.6 \citep[e.g.][]{Hummer1987}. For consistency with other surveys (as discussed by e.g. \citealt{HayesReview,Henry2015}), we assume $n_e \approx 350$ cm$^{-3}$ and $T=10^4$K, such that the intrinsic ratio between Ly$\alpha$ and H$\alpha$ is 8.7.
Therefore, we define the Ly$\alpha$ escape fraction as:
\begin{equation}
{\rm f}_{\rm esc} = \frac{f_{\rm Ly\alpha, obs}}{8.7 f_{\rm H\alpha , corrected}}
\end{equation}
In the presence of an AGN, the assumption of case B recombination is likely invalid because e.g. collisional ionisation might play a role due to shocks, leading to false estimates of the escape fraction. Among the sample of HAEs, we identify nine X-ray AGN in COSMOS using {\it Chandra} detections \citep{Elvis2009}, which are all spectroscopically confirmed to be at $z=2.23$ \citep{Civano2012}. Eight of these are significantly detected in NB392 imaging. We exclude these AGN from stacking analyses, but will keep them in our sample for studying individual sources.

Note that since we measure line-fluxes in 3$''$ apertures, f$_{\rm esc}$ is strictly speaking the escape fraction within a radius of 12 kpc. It is possible that the total escape fraction is higher, particularly in the presence of an extended low surface brightness halo due to resonant scattering (see also the discussion from a modellers point of view by \citealt{Zheng2010}).

\subsection{Corrections to measurements}
Although our matched NB survey requires less assumptions and uncertain conversions than escape fraction estimates based on UV or other emission-line measurements, we still need to take the following uncertainties/effects into account: 
\begin{enumerate}
\item interlopers in the H$\alpha$ sample (\S 3.4.1)
\item different filter transmissions (\S 3.4.2)
\item dust correction of the observed H$\alpha$ flux (\S 3.4.3)
\item {\sc [Nii]} contributing to the flux in the H$\alpha$ filter (\S 3.4.4)
\end{enumerate}

\subsubsection{Interlopers}
Our H$\alpha$ sample is selected using photometric redshifts and colour-colour techniques in a sample of emission line galaxies obtained from NB imaging (see \citealt{Sobral2013}). This means that galaxies with other emission lines than H$\alpha$ can contaminate the sample if the photometric redshift is wrongly assigned (for example if the galaxy has anomalous colours). Spectroscopic follow-up shows a 10\% interloper fraction, although this follow-up is so far limited to the brightest sources. These interlopers are either dusty low redshift ($z<1$) sources, such as Pa$\beta$ at $z=0.65$, or H$\beta$/{\sc[Oiii]} emitters ($z\sim3.2-3.3$, e.g. \citealt{Khostovan2015}). For the $z\sim3.3$ emitters, the NB392 would only measure noise, as the NB392 filter observes below the Lyman break for higher redshift galaxies and the flux for the low redshift interlopers is typically much fainter than the NB392 limit. The identified interlopers do not occupy a particular region in the parameter space of the sample of HAEs. There may be small dependences of contamination with galaxy properties, but no trends are seen for our limited follow-up, thus we assume a flat contamination. For stacking, we increase our observed NB392 flux by 10\% to account for these interlopers. For individual sources without NB392 detection, we are careful in our analysis as there is the risk of interlopers, even though the fraction is relatively small. 

\subsubsection{Filter transmissions}
While the NB392 and NB$_K$ filters are very well matched in terms of redshift coverage, the transmission at fixed redshift varies between H$\alpha$ and Ly$\alpha$. This means that the measured escape fraction is influenced by the particular redshift of the galaxy and resulting different filter transmissions for H$\alpha$ and Ly$\alpha$. Furthermore, systematic velocity offsets between Ly$\alpha$ and H$\alpha$ might increase this effect, as it has been found that Ly$\alpha$ is redshifted typically 200 (400) km s$^{-1}$ with respect to H$\alpha$ in Ly$\alpha$ (UV) selected galaxies \citep[e.g.][]{Steidel2010,McLinden2011,Kulas2012,Hashimoto2013,Erb2014,Shibuya2014,Song2014,Sobral2015CR7,Trainor2015}. We test the effect of the different filter transmissions and velocity shifts using a Monte Carlo simulation, similar to e.g. \cite{Nakajima2012}. We simulate 1,000,000 galaxies with redshifts between the limits of the NB$_K$ filter, and with a redshift probability distribution given by the NB$_K$ filter transmission (as our sample is H$\alpha$ selected). Then, we redshift the Ly$\alpha$ line w.r.t. H$\alpha$ with velocity shifts ranging from $0-800$ km s$^{-1}$, and fold it through the filter transmission in NB392. Finally, we compute the average relative H$\alpha$-Ly$\alpha$ transmission. For a zero velocity offset, the average transmission is 20\% higher for Ly$\alpha$ than H$\alpha$, because the NB392 filter is more top-hat like than the NB$_K$ filter. Increasing the velocity offset leads to an average lower Ly$\alpha$ transmission, as it is redshifted into lower transmission regions in the right wing of the filter. This effect is however very small, as it is constant up to a velocity shift of 400 km s$^{-1}$, and decreases to 11\% for 800 km s$^{-1}$. Because of this, we decrease the Ly$\alpha$-H$\alpha$ ratio of stacks and individual sources by 20\%. We add the 20\% uncertainty of this correction to the error on the escape fractions in quadrature. Spectroscopic follow-up is required to fully investigate the effect of velocity offsets on our measured escape fractions. 
\begin{table*}
\centering
\begin{tabular}{rrrrrrrr}
\hline
{\bf H$\alpha$ sample}  & {\bf Nr.} &$\langle f_{\rm H\alpha} \rangle$& $\langle f_{\rm Ly\alpha} \rangle$  &$\log_{10}{\left( {\rm M}_{\rm star} \right)}$ & \bf A$_{\rm H\alpha, Calzetti}$ & \bf A$_{\rm H\alpha, Garn\&Best}$ &{\bf f$_{\rm esc}$}  \\ 	
 & &[$10^{-16}$ erg s$^{-1}$ cm$^{-2}$] &[$10^{-16}$ erg s$^{-1}$ cm$^{-2}]$& [M$_{\odot}$]& [mag] & [mag]  & [\%] \\ \hline
 SFG with Ly$\alpha$ & 12 & 0.5$\pm0.3$ & 0.7$\pm0.5$ & 10.3$\pm0.8$& 0.83$\pm0.4$ & 1.11$\pm0.4$ &  10.8$\pm1.3$ \\  
 SFG no Ly$\alpha$ & 468 & 0.4$\pm0.3$ & $<0.7$ & 9.9$\pm0.7$ & 0.83$\pm0.5$ & 0.86$\pm0.4$ & $<20.1$ \\ 
 AGN with Ly$\alpha$ & 5 & 1.3$\pm0.4$& 3.6$\pm1.5$  & 10.8$\pm0.4$& 0.50$\pm0.3$ & 1.55$\pm0.3$& 12.8$\pm1.4$* \\
 SFGs for stacks & 265 & 0.4$\pm0.3$& 0.1$\pm0.01$  & 9.9$\pm0.7$& 1.0$\pm0.4$ & 0.86$\pm0.4$& 0.9$\pm0.1$ \\

\hline\end{tabular}
\caption{Numbers and median properties with 1$\sigma$ deviations of the sample of H$\alpha$-Ly$\alpha$ emitters with and without AGN. We also show the upper limits on the galaxies that are not detected in Ly$\alpha$, which is the comparison sample. Note that these are the median upper limits. H$\alpha$ and Ly$\alpha$ fluxes are the values observed in 3$''$ apertures. For completeness, we show the subsample of star-forming galaxies (SFGs) that we used for stacks (these are selected based on the depth of Ly$\alpha$ observations). The masses are derived from SED fitting (\citealt{Sobral2014}), which also gives H$\alpha$ attenuation based on the stellar extinction and the \citet{Calzetti2000} law (see \S 3.5). The H$\alpha$ attenuation from \citet{GarnBest2010} is based on a calibration between dust, SFR and mass. The total sample consists of 488 HAEs, with a stacked median escape fraction of 0.9$\pm0.1$\% (for 3$''$ diameter apertures), which is lower than the median escape fraction of individually detected source, because for individual sources we are observationally biased towards high escape fractions. 
*This escape fraction is likely wrong, as in AGN there is likely a departure from case B recombination due to shocks. We still show this for comparison, indicating that Ly$\alpha$ is typically bright for AGN. }
\label{tab:numbers}
\end{table*}

\subsubsection{Dust attenuation}
Even though the H$\alpha$ emission line is at red wavelengths compared to, for example, UV radiation, it is still affected by dust, such that we underestimate the intrinsic H$\alpha$ luminosity. Correcting for dust typically involves a number of uncertainties, such as the shape and normalisation of the attenuation curve, the difference between nebular and stellar extinction (e.g. \citealt{Reddy2015} and references therein) and the general uncertainties in SED fitting. For consistency with other surveys, we correct for extinction by applying a \cite{Calzetti2000} dust correction, using the estimated extinction, E$(B-V)_{\rm star}$, measurements from the best fit SED model from \cite{Sobral2014}. Note that we assume E$(B-V)_{\rm star}$ = E$(B-V)_{\rm gas}$, independent of galaxy property. Recent spectroscopic results at $z\sim2$ \citep[e.g.][]{Reddy2015} indicate that this is reasonable when averaged over the galaxy population, although there are indications that the nebular attenuation is higher than the stellar attenuation for galaxies with high SFR, particularly for galaxies with SFR $>50$ M$_{\odot}$ yr$^{-1}$. Therefore, if such a trend would be confirmed, our inferred relations between f$_{\rm esc}$ may be slightly affected. We discuss this when relevant in \S 6 and \S 8.2.1.

When stacking, we use the median dust correction of the sources included in the stacked sample, which is A$_{\rm H\alpha}$ = 1.0. This number has also been used for example by \cite{Sobral2013} in order to derive the cosmic star formation rate density, which agrees very well with independent measures. \cite{Ibar2013} showed this median attenuation also holds for a similar sample of HAEs at $z=1.47$ by using Herschel data.  

However, we also investigate how our results change when using the dust correction prescription from \cite{GarnBest2010}, which is a calibration between dust extinction and stellar mass based on a large sample of spectroscopically measured Balmer decrements in the local Universe. This relation between Balmer decrement and stellar mass is shown to hold up to at least $z\sim1.5$ \citep[e.g.][]{Sobral2012,Ibar2013}.

For individual sources, the two different dust corrections explored here can vary by up to a factor five, as seen in Table $\ref{tab:catalog}$. This results in large systemic errors which can only be addressed with follow-up spectroscopy to measure Balmer decrements. Throughout the paper, we add the error on the dust correction due to the error in SED fitting in quadrature to the error of the H$\alpha$ flux, but note that the systematic errors in the dust-correction are typically of a factor of two. 

\subsubsection{{\sc[Nii]} contamination}
Due to the broadness of the NB$_K$ filter used to measure H$\alpha$, the adjacent {\sc[Nii]} emission line doublet contributes to the observed line-flux. We correct for this contribution using the relation from \cite{Sobral2012}, who calibrated a relation between {\sc [Nii]}/({\sc [Nii]}+H$\alpha$) and EW$_{0,\, \rm H\alpha+[NII]}$ on SDSS galaxies. More recently, \cite{Sobral2015} found the relation to hold at least up to $z\sim1$. At $z=2.23$, we use this relation to infer a typical fraction of {\sc [Nii]}/({\sc [Nii]}+H$\alpha$) $= 0.17\pm0.08$, which is consistent with spectroscopic follow-up at $z\sim2$ \citep{Swinbank2012,Sanders2015}. We have checked that our observed trends between f$_{\rm esc}$ and galaxy properties do not qualitatively depend on this correction - if we apply the median correction to all sources, the results are the same within the error bars. We add 10 \% of the correction to the error in quadrature. For stacks, we measure the EW$_{0,\, \rm H\alpha+[NII]}$ and apply the corresponding correction, which is consistent with the median correction mentioned here, and we also add 10\% of the correction to the error in quadrature.

\subsection{Definitions of galaxy properties}
We compare f$_{\rm esc}$ with a range of galaxy properties, defined here. SFRs are computed from H$\alpha$ luminosity, assuming a luminosity distance of $17746$ Mpc (corresponding to $z=2.23$ with our cosmological parameters) and the conversion using a \cite{Chabrier2003} IMF: 
\begin{equation}
{\rm SFR(H}\alpha{\rm )/(M}_{\odot}{\rm yr^{-1})} = 4.4\times10^{-42}\, {\rm L(H}\alpha{\rm )/(erg\, s}^{-1}{\rm )} 
\end{equation}
where L(H$\alpha$) is the dust-corrected H$\alpha$ luminosity and SFR(H$\alpha$) the SFR. 

Stellar masses and extinctions (E$(B-V)$) are obtained through SED fitting as described in \cite{Sobral2014}. In short, the far UV to mid-infrared photometry is fitted with \cite{BC2003} based SED templates, a \cite{Chabrier2003} IMF, exponentially declining star formation histories, dust attenuation as described by \cite{Calzetti2000} and a metallicity ranging from $Z= 0.0001 - 0.05$. While we use a mass defined as the median mass of all fitted models within 1$\sigma$ of the best fit, we use the E$(B-V)$ value of the best fitted model. The errors on stellar masses and extinctions are computed as the 1$\sigma$ variation in the fitted values from SED models that have a $\chi^2$ within 1$\sigma$ of the best fitted model. For stellar mass, these errors range from 0.2 dex for the lowest masses to 0.1 dex for the highest masses. The typical uncertainty on the extinction ranges from 0.12 at E$(B-V)\approx0.1$, to 0.05 at E$(B-V)\approx0.3$.

The UV slope $\beta$ (which is a tracer for dust content, stellar populations and escape of continuum ionising photons, e.g. \citealt{Dunlop2012}) is calculated using photometry from the observed $g^+ - R$ colours. These bands were chosen such that there is minimal contribution from Ly$\alpha$ to the $g^+$ band (the transmission at the corresponding wavelength is $<5$\%), and such that we measure the slope at a rest-frame UV wavelength of $\sim 1500${\AA}. We also chose to derive the slope from observed colours in stead of using the SED fit, as otherwise there might be biases (e.g. the SED based extinction correction is related to the UV slope). The error in $\beta$ due to measurement errors in $g^+$ and $R$ ranges from typically 0.5 at $\beta = -2.3$ to 0.3 at $\beta > 0$.

\begin{figure}
	\includegraphics[width=8.5cm]{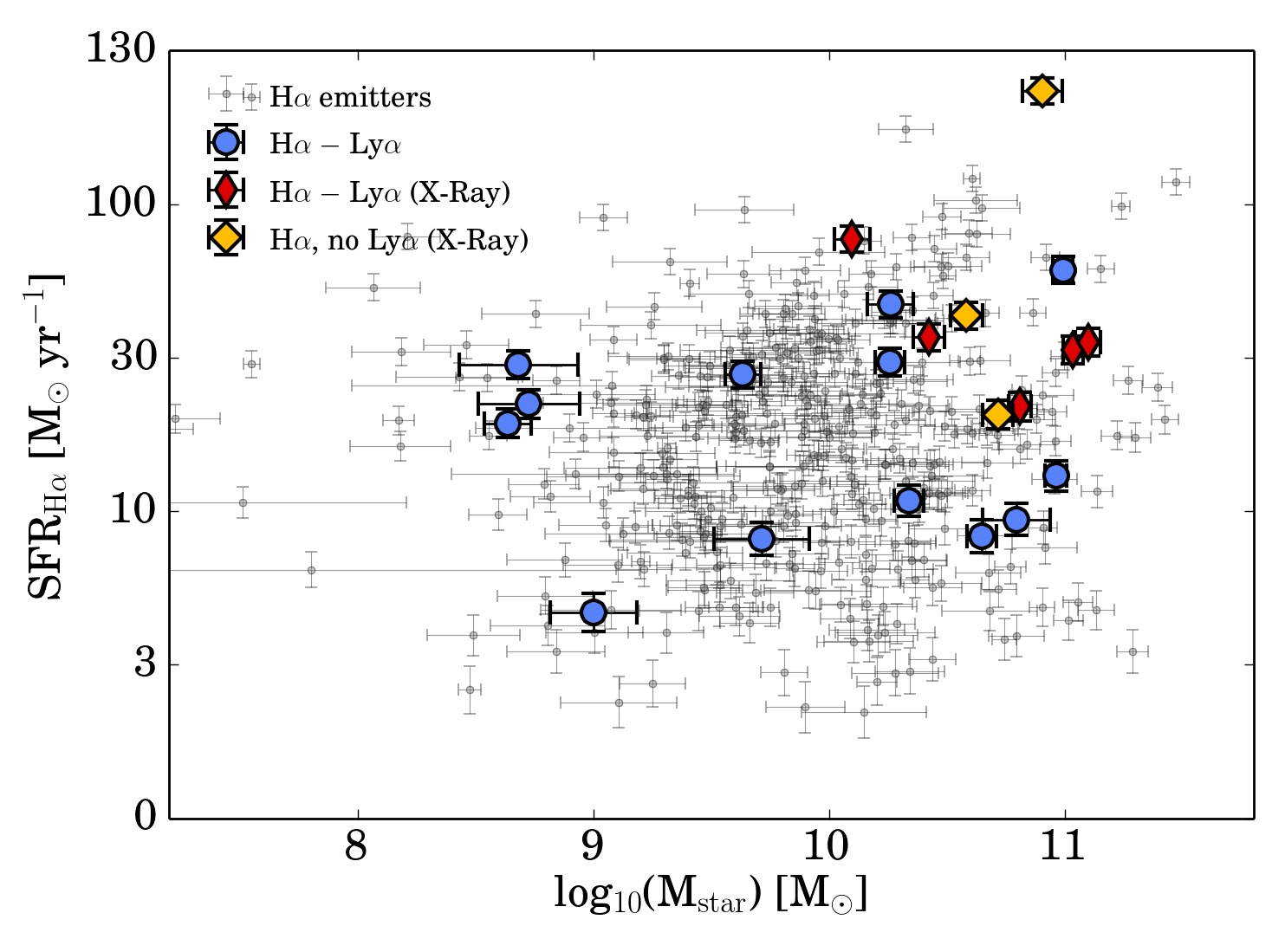}
    \caption{SFR(H$\alpha$) versus stellar mass for the observed H$\alpha$ emitters. We obtained the SFR from dust corrected H$\alpha$ and stellar mass from SED fitting (see \S 3.5). We show the position of sources with and without Ly$\alpha$, AGN with Ly$\alpha$ and AGN without Ly$\alpha$. There is no obvious difference in the SFRs or stellar masses between sources with or without Ly$\alpha$. Since it is easier to observe Ly$\alpha$ for galaxies with higher SFR (at a given  f$_{\rm esc}$), this already indicates that f$_{\rm esc}$ is higher for galaxies with low SFR. Note that the SFR for the AGN is likely to be overestimated as AGN activity also contributes to the H$\alpha$ flux.}
    \label{fig:sfr_mstar}
\end{figure}

\section{Stacking method}
In order to reach deeper Ly$\alpha$ line-fluxes, we use stacking methods to combine observations of our full sample of observed galaxies, such that the exposure time is effectively increased by a factor of $\sim 400$. This however involves some complications and assumptions. For example, we will use the median stacked value, rather than the mean stacked value, such that our results are not biased towards bright outliers. However, our results will still be biased towards the most numerous kind of sources in our sample. Stacking also assumes that all sources are part of a single population with similar properties - which may not always be the case, as indicated by the results in the previous section. 

We divide our sample in subsets of various physical properties in \S 6 and study how these stacks compare with the results from individual galaxies. We discuss the effect of varying apertures in \S 7. The errors of the measured fluxes and resulting escape fractions in stacks are estimated using the jackknife method. The errors due to differences in the PSF of the NB and broadband are added as a function of aperture radius (see  \S 4.1). We add all other sources of systematic error (see \S 3.4) in quadrature.

We obtain stacked measurements by median combining the counts in $1'\times1'$ thumbnails in $U$, $B$, NB392, NB$_K$ and $K$ bands of the H$\alpha$ emitters covered in our INT observations (see Fig. $\ref{fig:RADEC}$). From the stacked thumbnails (as for example shown in Fig. $\ref{fig:thumbnails}$), we measure photometry in various apertures at the central position (defined by the position of the NB$_K$ detection. Note that our typical astrometric errors are of the order $\sim0.1''$, corresponding to $\sim1$ kpc). With this photometry, we obtain line-fluxes for both Ly$\alpha$ and H$\alpha$. The Ly$\alpha$ flux is corrected using $U-B$ colours, and we account for the {\sc[Nii]} contribution to the NB$_K$ flux using the relation with EW from \cite{Sobral2012} (see \S 3.4.4). We also add the error due to differences in the PSF of $U$ and NB392 to the error of the Ly$\alpha$ flux (see \S 4.1). We apply the median dust correction of the H$\alpha$ emitters, which is roughly similar for using the Calzetti or Garn \& Best method: A$_{\rm H\alpha}$ = 1.0 or A$_{\rm H\alpha}$ = 0.86, respectively. For our full sample of 488 H$\alpha$ emitters, we observe a median stacked Ly$\alpha$ line-flux of $3.5\pm0.3\times10^{-18}$ erg s$^{-1}$ cm$^{-2}$, and an escape fraction of 0.3$\pm0.06$ \% in 3$''$ apertures, corresponding to a radial distance to the centre of $\sim 20$ kpc. The significance of these detections are discussed in \S 7. 

The depth of our NB392 observations is inhomogeneous over the full fields (see Fig. $\ref{fig:RADEC}$). We therefore study the effect of limiting our sample based on the depth of the NB392 observations. We find that the photometric errors on the stacked NB392 image are minimised when we only include sources for which the local 3$\sigma$ depth is at least 24.1 AB magnitude, which corresponds to the inclusion of 265 out of the 488 sources. For the remainder of this section, we only include sources which are among these 265. The median SFRs, stellar masses, and dust attenuations of this sample are similar to the average properties of the full sample (see Table $\ref{tab:numbers}$). The 3$\sigma$ depth of the NB392 stack of these 265 sources is 27.2 AB magnitude. In the case of a pure line and no continuum contributing to the NB392 flux, this corresponds to a limiting line-flux of $\sim 5\times10^{-18}$ erg s$^{-1}$ cm$^{-2}$.

\begin{figure}
	\includegraphics[width=8.6cm]{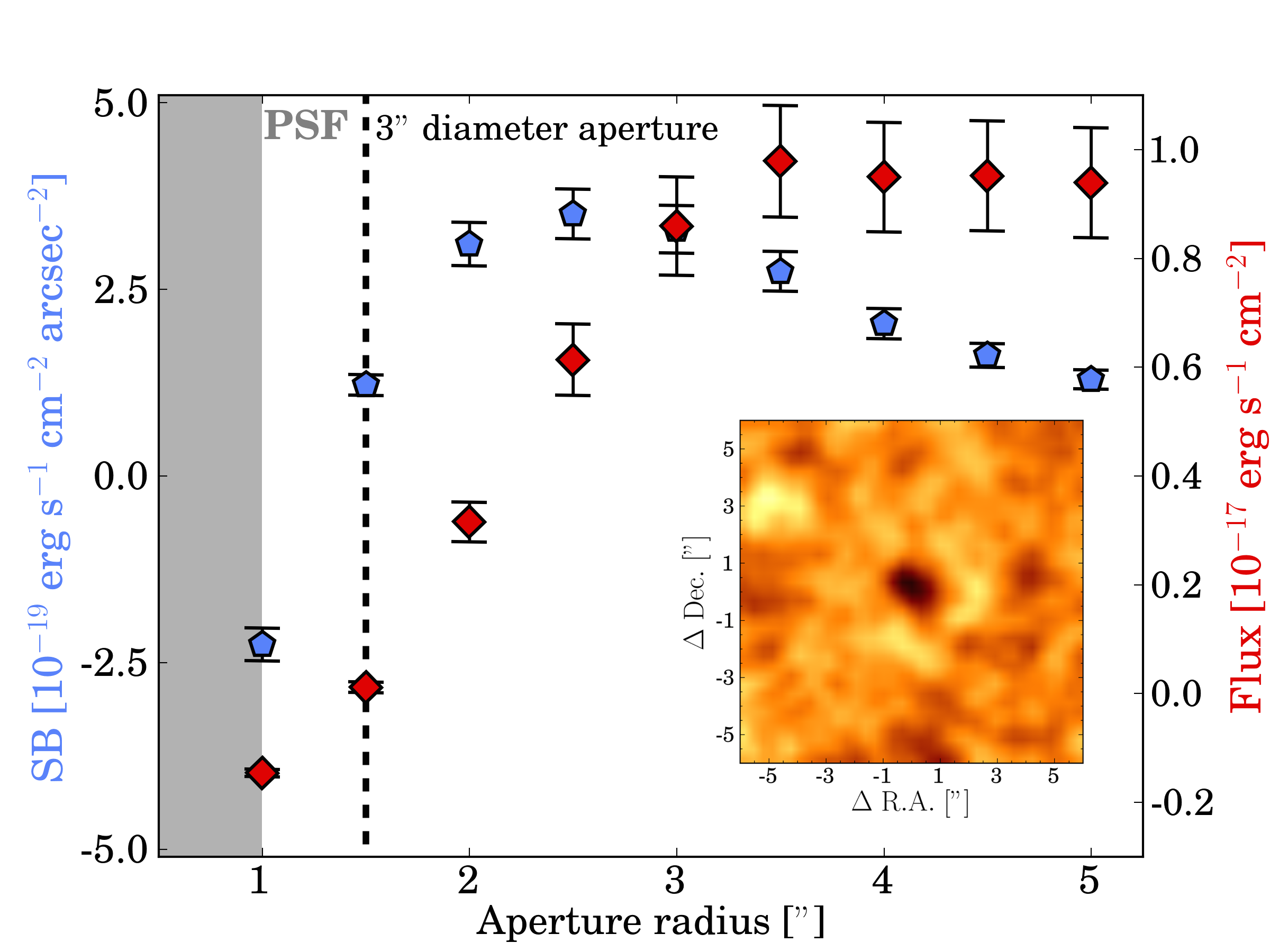}
    \caption{Surface brightness profile (blue) and integrated flux (red) of the stack of the reference sample, which should have a zero line-flux at a 1.5$''$ aperture radius by construction. We indicate the 1.5$''$ aperture radius with a dashed black line. The inset figure shows the 2D image obtained by subtracting the BB from the NB. We find a small residual signal which has central absorption and peak at a radial distance of 2.5$''$. We attribute this signal due to differences in the PSF of the NB and the broadband. In the remainder of this paper, we subtract this signal from the surface brightness of individual sources and from stacks and add this subtraction to the error of the total flux in quadrature. The typical signal measured for individually detected HAEs is typically 10-100 times higher than the signal due to the PSF differences, but it is of the same order of magnitude as stacked measurements.} 
    \label{fig:error}
\end{figure}

\subsection{Empirical evaluation of different PSF shapes}
The NB and broadband observations are taken with different telescopes, cameras and at different observing sites and under different conditions. Therefore, even though we match the PSF-FWHM of all images, the actual shape of the PSF might vary between NB and broadband. This might artificially influence the surface brightness profile of line-emission estimated from the difference between the two bands. This becomes particularly important when we study stacked images of over 300 sources, where errors on the percent level might dominate the measured signal.

We empirically evaluate the differences in NB and broadband PSF by performing the following sanity check: we first select line-emitters in NB$_K$ imaging, which: i) are {\it not} selected as H$\alpha$ emitters, ii) are  {\it not} selected as higher redshift line-emitters, or iii) do  {\it not} have a photometric redshift $>1$ (from \citealt{Ilbert2009}). With this sample, we ensure that the NB392 photometry should measure (relatively flat) continuum by removing a handful of sources with an emission line in NB392. This leaves us with 245 sources, which have a similar NB$_K$ magnitude distribution as the HAE sample.
The $U$, $B$ and NB392 images are stacked in the exact same way as we treat the HAEs. This is used to measure the resulting line-flux and surface brightness profile of the stack in the NB392 band (we estimate the continuum from $U$ and $B$). 
Although the NB392 was photometrically calibrated to the $U$ band in 3$''$ diameter apertures, we detect a small residual signal with a typical surface brightness profile of central absorption (with surface brightness $\sim -2\times10^{-19}$ erg s$^{-1}$ cm$^{-2}$ arcsec$^{-2}$, see Fig. $\ref{fig:error}$), and peaking at a radial distance of 2.5$''$ (with a surface brightness of $\sim 4\times10^{-19}$ erg s$^{-1}$ cm$^{-2}$ arcsec$^{-2}$). We note that at the radial aperture of 1.5$''$, which we used for our calibration, the integrated flux signal is consistent with zero, see Fig. $\ref{fig:error}$. Corrections therefore only need to be applied for other aperture radii and surface brightness profiles. For individually detected Ly$\alpha$ sources, the residual signal at the $1-10$\% flux level, but for stacks, it can be more important. The origin of this residual signal is likely because of differences in the inner part of the PSF, similarly as those reported by e.g. \cite{Momose2014}. The uncertainty in our astrometry is of the order of 0.1$''$ and therefore likely less important.

We thus conclude that the differences in PSF shapes of broadband and narrow-band have a small effect on stacked measurements, but we still take it into account by correcting all surface brightness profiles and any aperture measurements at values other than 3$''$. We add the residual flux to the error of the total flux in quadrature. 

\begin{table*}
\centering
\begin{tabular}{lllrrrrrrrr}
\hline
ID & R.A. & Dec. & M$_{\rm star}$ & f$_{\rm Ly\alpha}$ & EW$_{0,\rm Ly\alpha}$ & f$_{\rm H\alpha}$ & A$_{\rm H\alpha, C}$ &A$_{\rm H\alpha, GB}$ & f$_{\rm esc}$ & Note$^3$ \\ 
 & (J2000) & (J2000) & log$_{10}$(M$_{\odot}$) & erg/s/cm$^{2}$ & {\AA} & erg/s/cm$^{2}$ & SED$^1$ & GB$^2$ & \% & \\ 
 & & &  & $\times10^{-16}$ & & $\times10^{-16}$ & & & & \\ \hline
1057 & 10:00:39.6 & +02:02:41.2 & 10.6$^{+0.1}_{-0.1}$ & 0.46 &82 & 0.42  & 0.17 & 1.43 & 10.8 $\pm$ 1.4 & \\
1073 & 10:00:44.2 & +02:02:06.9 & 11.1$^{+0.1}_{-0.1}$ & 0.88 & 55 & 1.24  & 0.50 & 1.77 & 5.1 $\pm$ 0.6 & 1\\ 
1139 & 10:00:55.4 & +01:59:55.4 & 10.8$^{+0.1}_{-0.1}$ & 4.55  & 63 & 1.03  & 0.17 & 1.56 & 43.7 $\pm$ 5.1 & 1, 2, 3\\
1993 & 10:02:08.7 & +02:21:19.9 & 8.6$^{+0.1}_{-0.1}$ & 1.14  & 68 & 0.26  & 1.49 & 0.29 & 12.8 $\pm$ 1.5 & 3\\
2600 & 10:00:07.6 & +02:00:13.2 & 8.7$^{+0.2}_{-0.2}$ & 1.32  & 80 & 0.53  & 1.00 & 0.29 & 11.4 $\pm$ 1.3 & 3\\
2741 & 10:01:57.9 & +01:54:36.9 & 10.1$^{+0.1}_{-0.1}$ & 3.55  & 14 & 2.24  & 0.67 & 0.99 & 9.9 $\pm$ 1.0 & 1\\
4032 & 10:00:51.1 & +02:41:16.9 & 11.0$^{+0.1}_{-0.1}$ & 0.46  & 28 & 0.34  & 0.83 & 1.67 & 7.3 $\pm$ 1.1 & \\
4427 & 10:01:19.4 & +02:07:32.6 & 8.7$^{+0.2}_{-0.4}$ & 0.57  & 12 & 0.58  & 1.16 & 0.29 & 3.8 $\pm$ 0.6 & \\
4459 & 10:01:43.3 & +02:11:15.7 & 10.3$^{+0.1}_{-0.1}$ & 1.41  & 93 & 0.41  & 0.50 & 1.18 & 24.8 $\pm$ 3.1 & \\
4861 & 10:00:03.3 & +02:11:04.4 & 9.0$^{+0.2}_{-0.2}$ & 0.31  & 16 & 0.28  & 0.0 & 0.34 & 12.4 $\pm$ 2.3 & \\ 
5583 & 10:01:59.6 & +02:39:32.7 & 10.8$^{+0.2}_{-0.1}$ & 1.73  & 244 & 0.41  & 0.50 & 1.54 & 30.4 $\pm$ 3.8 & \\
5847 & 10:01:12.2 & +02:53:25.9 & 10.3$^{+0.1}_{-0.1}$ & 0.70  & 60 & 1.03  & 0.50 & 1.11 & 4.9 $\pm$ 0.5 & \\ 
7232 & 10:01:05.4 & +01:46:11.6 & 10.3$^{+0.1}_{-0.1}$ & 0.42  & 15 & 0.74  & 1.33 & 1.11 & 1.9 $\pm$ 0.3 & \\
7693 & 09:59:49.6 & +01:50:24.7 & 9.6$^{+0.1}_{-0.1}$ & 0.75  & 10 & 0.84  & 0.67 & 0.65 & 5.5 $\pm$ 0.8 & \\
7801 & 10:02:08.6 & +01:45:53.6 & 10.4$^{+0.1}_{-0.1}$ & 2.38  & 5 & 1.35  & 0.50 & 1.24 & 12.8 $\pm$ 1.4 & 1\\
9274 & 10:00:26.7 & +01:58:23.0 & 11.0$^{+0.1}_{-0.1}$ & 5.06  & 142 & 1.54  & 0.13 & 1.72 & 32.4 $\pm$ 3.6 & 1, 2 \\
9630 & 10:02:31.3 & +01:58:16.5 & 9.7$^{+0.2}_{-0.2}$ & 0.61  & 29 & 0.44  & 0.0 & 0.70 & 16.1 $\pm$ 2.2 & \\
\hline
\end{tabular}
 \caption{Properties of H$\alpha$ emitters at $z=2.23$ detected at $3\sigma$ in NB392 imaging and having a Ly$\alpha$ emission line. The IDs correspond to the last digits of the full HiZELS IDs, which can be retrieved by placing ``HiZELS-COSMOS-NBK-DTC-S12B-" in front of them. The coordinates are measured at the peak of NB$_K$ (rest-frame $R$+H$\alpha$) emission. The observed line-fluxes, EW, dust extinction and escape fraction are measured with 3$''$ apertures. The escape fraction (f$_{\rm esc}$) is computed under the assumptions explained in \S 3.4, thus using A$_{\rm H\alpha, C}$. $^1$: SED: Dust correction using SED fitted E$(B-V)$ and a \citet{Calzetti2000} law. $^2$: GB: \citet{GarnBest2010} dust correction based on stellar mass. $^3$: codes in this column correspond to: 1: X-Ray AGN, 2:  {\sc [Oii]} line detected in NB$_J$, 3: {\sc [Oiii]} line detected in NB$_H$. }
 \label{tab:catalog}
\end{table*}

\begin{figure}
	\includegraphics[width=\columnwidth]{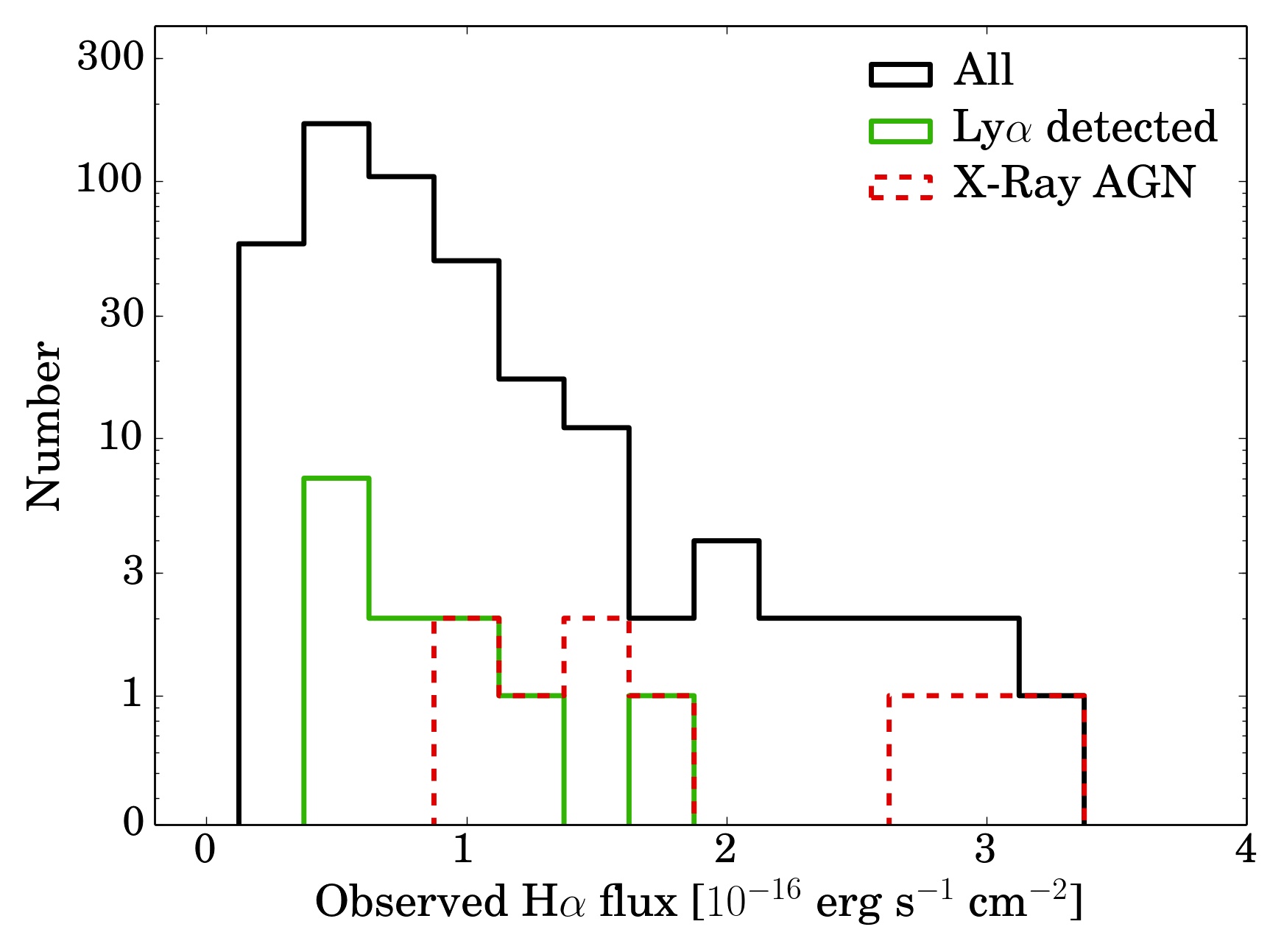}
    \caption{Histogram of H$\alpha$ fluxes in our galaxy sample at $z=2.23$. AGN are typically found among the brightest H$\alpha$ emitters, and are also typically detected in NB392, such that they are either bright in the UV continuum or Ly$\alpha$ or both. We also show the distribution of Ly$\alpha$ detected star-forming galaxies. It can be seen that not all brightest H$\alpha$ emitters have been detected, indicating very low escape fractions or interlopers. On the other hand, some very faint H$\alpha$ emitters are still detected in Ly$\alpha$, indicative of high Ly$\alpha$ escape fractions.}
    \label{fig:histogram}
\end{figure}

\section{Direct measurements for individual galaxies} 
We directly detect ($>3\sigma$) 43 out of our 488 HAEs in the NB392 imaging, which is a combination of UV continuum and Ly$\alpha$ line (see Table $\ref{tab:numbers}$). The 3$\sigma$ limit corresponds to limiting Ly$\alpha$ fluxes ranging from $3.8-7.4 \times10^{-17}$ erg s$^{-1}$ cm$^{-2}$ (assuming the typical continuum level of 0.23 $\mu$Jy, $\sim 25.5$ AB magnitude in the $U$ band). Out of these robust detections, 17 show a significant Ly$\alpha$ line detection (excess significance $\Sigma > 2$), all in COSMOS. The properties of these sources and their IDs from the HiZELS catalog \citep{Sobral2013} are listed in Table $\ref{tab:catalog}$. The other 26 robust NB392 detections are H$\alpha$ emitters with strong upper limits on their Ly$\alpha$ flux, as we have detected the UV continuum in the NB392 filter.\footnote{Note that these sources are unlikely higher-redshift interlopers, as the NB392 filter is below the Lyman break at $z>3$, such that a NB392 detection rules out that the source is a $z\sim3.3$ H$\beta$/{\sc[Oiii]} or $z=4.7$ {\sc[Oii]} emitter \citep[e.g.][]{Khostovan2015}. Moreover, 13 of these have detected {\sc [Oiii]}, {\sc [Oii]} or both lines.} 

Five of the dual emitters are matched (within 3$''$) with an X-ray detection from {\it Chandra}. From spectroscopy with IMACS and from $z$COSMOS \citep{Lilly2009}, these are all classed as BL-AGN. These AGN are among the brightest and most massive H$\alpha$ emitters (Fig. $\ref{fig:histogram}$): all have stellar masses above $10^{10.5}$ M$_{\odot}$ (Fig. $\ref{fig:sfr_mstar}$), and the fraction of BL-AGN is consistent with the results from \cite{Brizels}. The ISM conditions surrounding the AGN might lead to other ionising mechanisms than case B photo-ionisation, such as shocks, making it more challenging to measure the Ly$\alpha$ escape fraction as the intrinsic H$\alpha$-Ly$\alpha$ changes. 

Fig. $\ref{fig:histogram}$ shows the distribution of H$\alpha$ fluxes of our observed sample, of the AGN and also of the star-forming sources directly detected in Ly$\alpha$. Whether a source is detected in Ly$\alpha$ does not clearly correlate with H$\alpha$ flux, such that even very faint HAEs are detected. These very faint sources generally have high f$_{\rm esc}$, although we note that it is possible that these sources are detected at a redshift where the transmission in the H$\alpha$ filter is low. In the remainder of the paper, we will use the sample of 17 dual H$\alpha$-Ly$\alpha$ emitters for direct measurements of f$_{\rm esc}$, and use upper limits for the other 471 HAEs. 

\begin{figure*}
	\includegraphics[width=18cm]{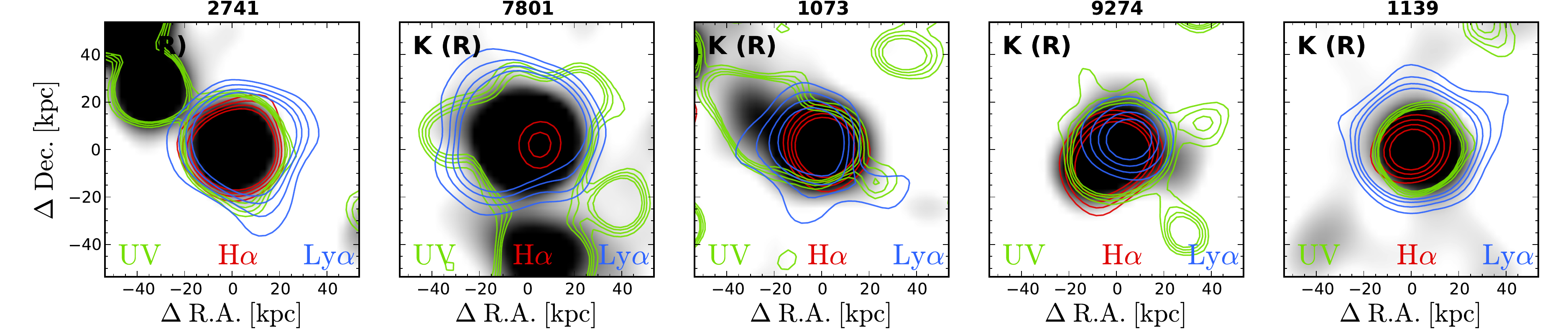}
        \includegraphics[width=18cm]{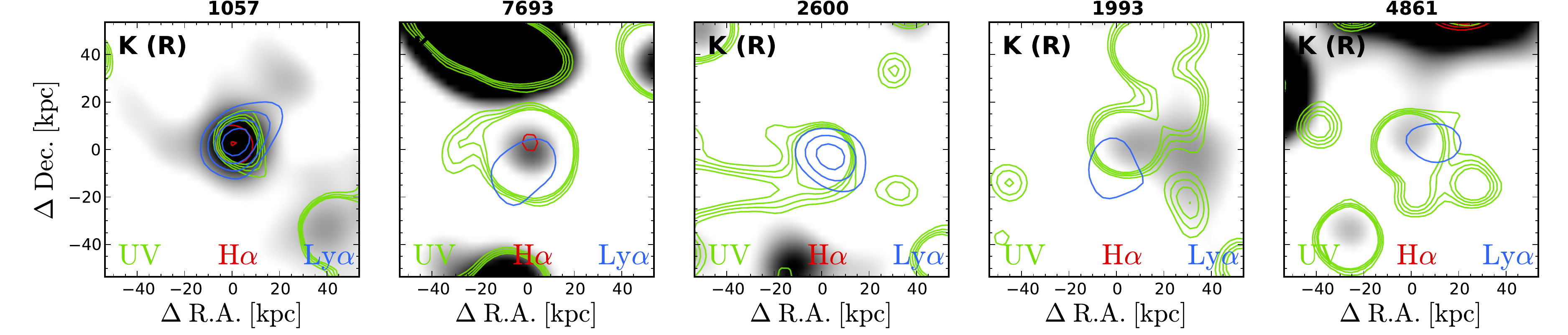}
    \caption{Ly$\alpha$ morphologies for the ten HAEs with most significant Ly$\alpha$ detections. All images (including {\it HST} ACS F814W) were smoothed to the PSF of the NB392 image and are centred on the position of peak H$\alpha$ emission. The {\it top} row shows AGN, the {\it bottom} row SFGs. The sources are ordered by H$\alpha$ flux, decreasing from left to right. Each panel shows a 100$\times$100 kpc $K$ band (which traces rest-frame $R$, and thus roughly stellar mass) thumbnail with contours of rest-frame Ly$\alpha$ (blue), UV (green, from ACS F814W) and H$\alpha$ (red). The Ly$\alpha$ image was obtained by subtracting the continuum PSF matched $U$ band from the NB392 image and correcting using the differential PSF image. The contours show 3, 4, 5 and 6 $\sigma$ levels in each respective filter. For Ly$\alpha$, the 3$\sigma$ contour corresponds to a surface brightness ranging from $1.8-17$ (median 2.0)$\times10^{-18}$ erg s$^{-1}$ cm$^{-2}$ arcsec$^{-2}$. The UV data is typically 3 magnitudes deeper than the Ly$\alpha$ data. Among the AGN, IDs 7801 and 1139 show evidence for extended Ly$\alpha$ emission. For the SFGs there is little convincing evidence for extended emission at the reached surface brightness limit. The Ly$\alpha$ emission of source 1993 appears to be offset from the peak UV emission. The faintest H$\alpha$ emitters have no 3$\sigma$ contour because of smoothing with the PSF of the Ly$\alpha$ image.}
    \label{fig:dual_detections}
\end{figure*}

\subsection{Ly$\alpha$ properties of dual Ly$\alpha$-H$\alpha$ emitters}
After excluding the X-ray AGN, we find 12 robust dual Ly$\alpha$-H$\alpha$ emitters, which would translate in 2.5\% of our star-forming galaxies being detected in Ly$\alpha$ down to our flux limit. However, if we only select the subset of HAEs with the deepest Ly$\alpha$ observations (194 star-forming galaxies, of which 8 have Ly$\alpha$), we find a fraction of 4.1\%. This is comparable to \citealt{Oteo2015}, who found a fraction of 4.5\% and lower than the 10.9 \% of \citealt{Hayes2010}, whose Ly$\alpha$ observations are a factor six deeper and H$\alpha$ sample consists of fainter sources (by a factor of seven), but covers a volume which is $\sim80$ times smaller than our survey. 

Comparing our 12 H$\alpha$-Ly$\alpha$ emitters to HAEs without Ly$\alpha$, we find that there are no clear differences in the SFR-M$_{\rm star}$ plane (Fig. $\ref{fig:sfr_mstar}$). Since it is easier to observe Ly$\alpha$ for galaxies with higher SFR (at a given f$_{\rm esc}$), this already indicates that f$_{\rm esc}$ is higher for galaxies with low SFR. We show the stellar masses, observed H$\alpha$ and Ly$\alpha$ fluxes and dust attenuations of the Ly$\alpha$ detected sources in Table $\ref{tab:catalog}$.

The median, dust-corrected H$\alpha$ luminosity for the 12 SFGs with Ly$\alpha$ is $4.5\times10^{42}$ erg s$^{-1}$(corresponding to a SFR of 20 M$_{\odot}$ yr$^{-1}$), and median stellar mass of $1.8\times10^{10}$ M$_{\odot}$, such that they have specific SFRs which are typical to the sample of HAEs, see Fig. $\ref{fig:sfr_mstar}$. The median Ly$\alpha$ luminosity is $2.8\times10^{42}$ erg s$^{-1}$ and the Ly$\alpha$ escape fraction for these galaxies ranges from $1.9\pm0.3$ to $30\pm3.8$ \% (although we note this does not take the uncertainty due to the filter transmission profiles into account, nor a statistical correction). The EW$_0$(Ly$\alpha$) ranges from 10 to 244 {\AA}. If we apply an EW$_0$(Ly$\alpha$) cut of $>25$ {\AA} (similar to the selection of LAEs at high redshift, e.g. \citealt{Matthee2014} and references therein), 7 out of 194 star-forming HAEs with deepest Ly$\alpha$ observations are recovered as LAEs, with luminosities $\sim 2-8 \times10^{42}$ erg s$^{-1}$.

\subsubsection{Morphology}
In Fig. $\ref{fig:dual_detections}$, we compare the Ly$\alpha$ surface brightness with rest-frame UV, $R$ and H$\alpha$ from {\it HST} ACS F814W \citep{Koekemoer2007}, $K$ and NB$_K$ (continuum corrected with $K$) respectively. In order to be comparable, the PSF of the {\it HST} images is matched to that of the NB392 imaging on the INT, using a convolution with a gaussian kernel. As the PSF of our INT imaging is $1.8-2.0$", this is a major limitation. However, for the most significantly detected sources (in Ly$\alpha$), it is still possible to study differences qualitatively. Even though there is ground based $I$ band data available, we use {\it HST} data because those are deeper.

The sources with IDs 2741, 7801 1073, 9274 and 1139 (see Table $\ref{tab:catalog}$ for more information), shown in the first row are all AGN with mostly symmetrical Ly$\alpha$ morphology. Compared to the UV, IDs 7801 and 1139 show evidence for extended Ly$\alpha$ emission, while this is more evident when Ly$\alpha$ is compared to H$\alpha$. Note that the Ly$\alpha$ image is typically the shallowest, and that the outer contours of Ly$\alpha$ therefore typically represent a higher fraction of the peak flux than the UV contours.

The sources in the second row are undetected in the X-ray, and therefore classed as SFGs. 1057 is relatively massive (M$_{\rm star} = 10^{10.6}$ M$_{\odot}$), and has f$_{\rm esc}$ of $10.8\pm1.6$\%. 7693 has an intermediate mass of $\sim10^{9.5}$ M$_{\odot}$ and escape fraction of $5.5\pm0.8$\%. From comparison with the H$\alpha$ image, it can be seen that Ly$\alpha$ preferentially escapes offset to the south from the galaxy centre, which might be indicative of an outflow.

The sources with IDs 2600, 1993 and 4861, in the last three columns of Fig. $\ref{fig:dual_detections}$ are the faintest HAEs for which we detect Ly$\alpha$ directly, such that there are no 3$\sigma$ H$\alpha$ contours due to smoothing the image with the PSF of the Ly$\alpha$ image. These HAEs have (possibly) little dust, blue UV slopes and Ly$\alpha$ EW$_0$ $>30$ {\AA}, such that they would be selected as Ly$\alpha$ emitters. The masses, SFRs and blue UV slopes are consistent with results from typical LAEs \citep[e.g.][]{Nilsson2009,Ono2010}, and similar to simulated LAEs \citep[e.g.][]{Garel2012,Garel2015}. The Ly$\alpha$ emission for ID 1993 and 4861 appears to be offset from the peak UV emission. This indicates that slit spectroscopy of UV or H$\alpha$ selected galaxies might miss significant parts of Ly$\alpha$. Note that we look at the stacked UV, Ly$\alpha$, H$\alpha$ and morphologies of these 12 SFGs and the full sample of SFGs in \S 7.

\begin{figure*}
\begin{tabular}{cc}
\includegraphics[width=8.6cm]{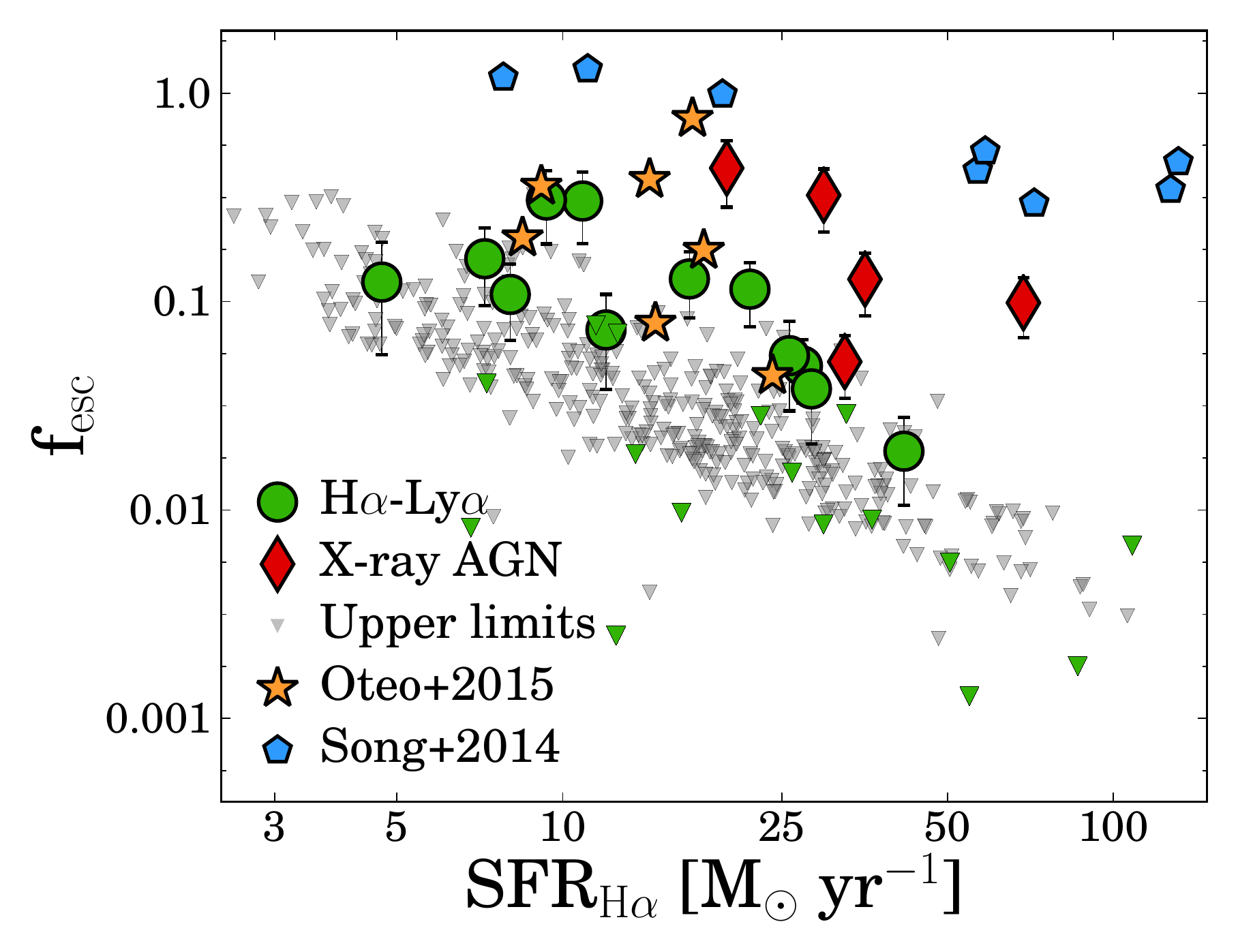}&

\includegraphics[width=8.6cm]{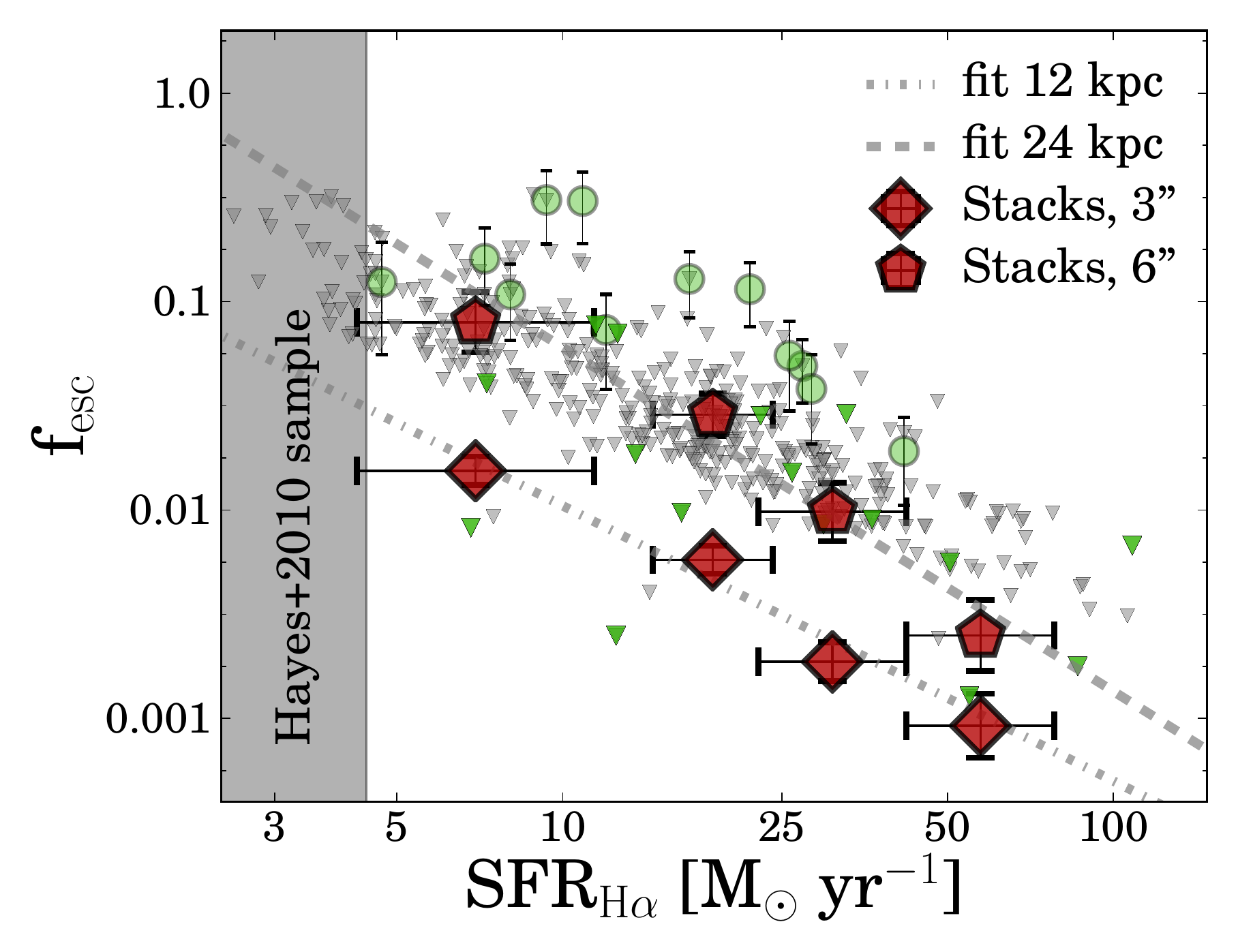}\\

\includegraphics[width=8.6cm]{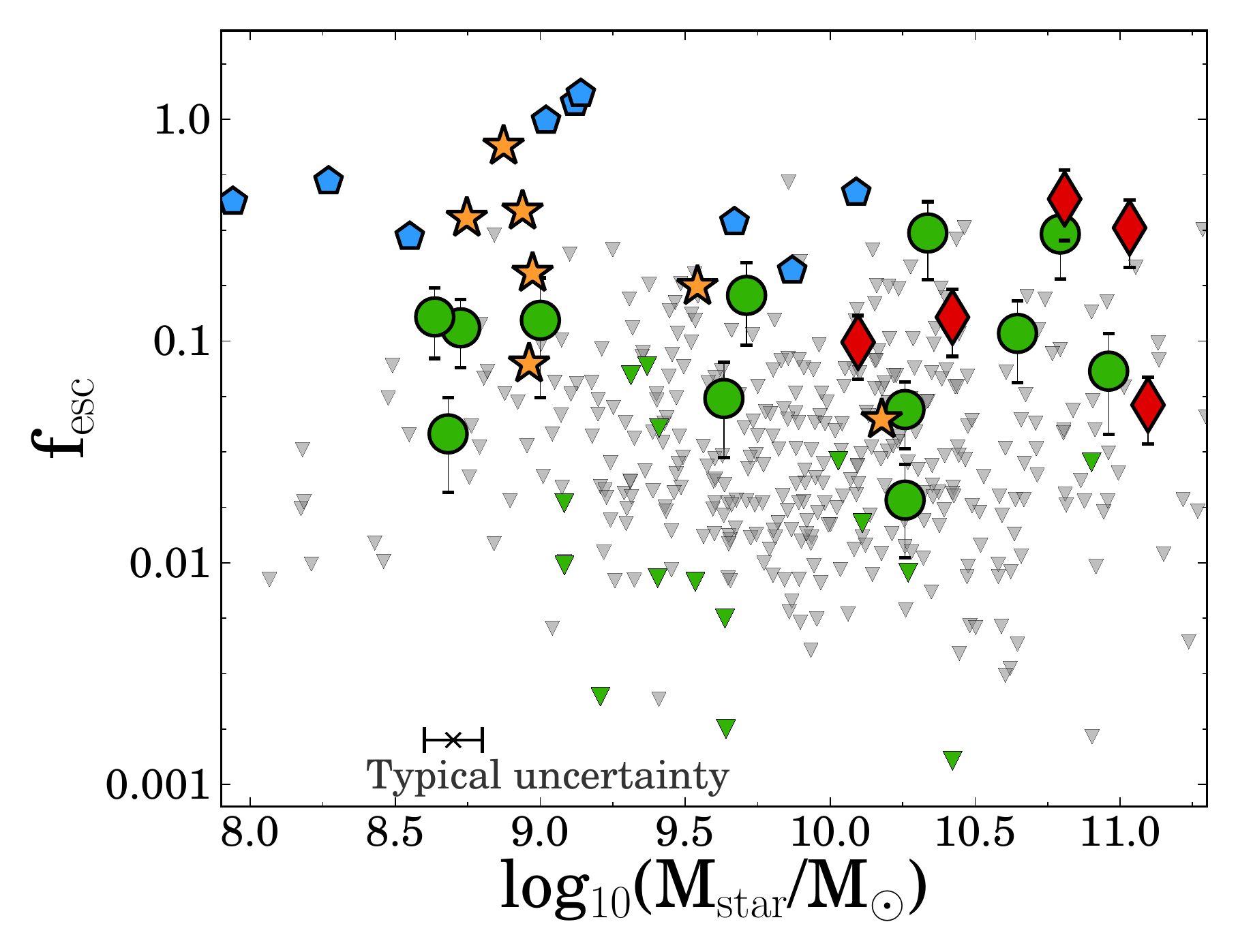}&

\includegraphics[width=8.6cm]{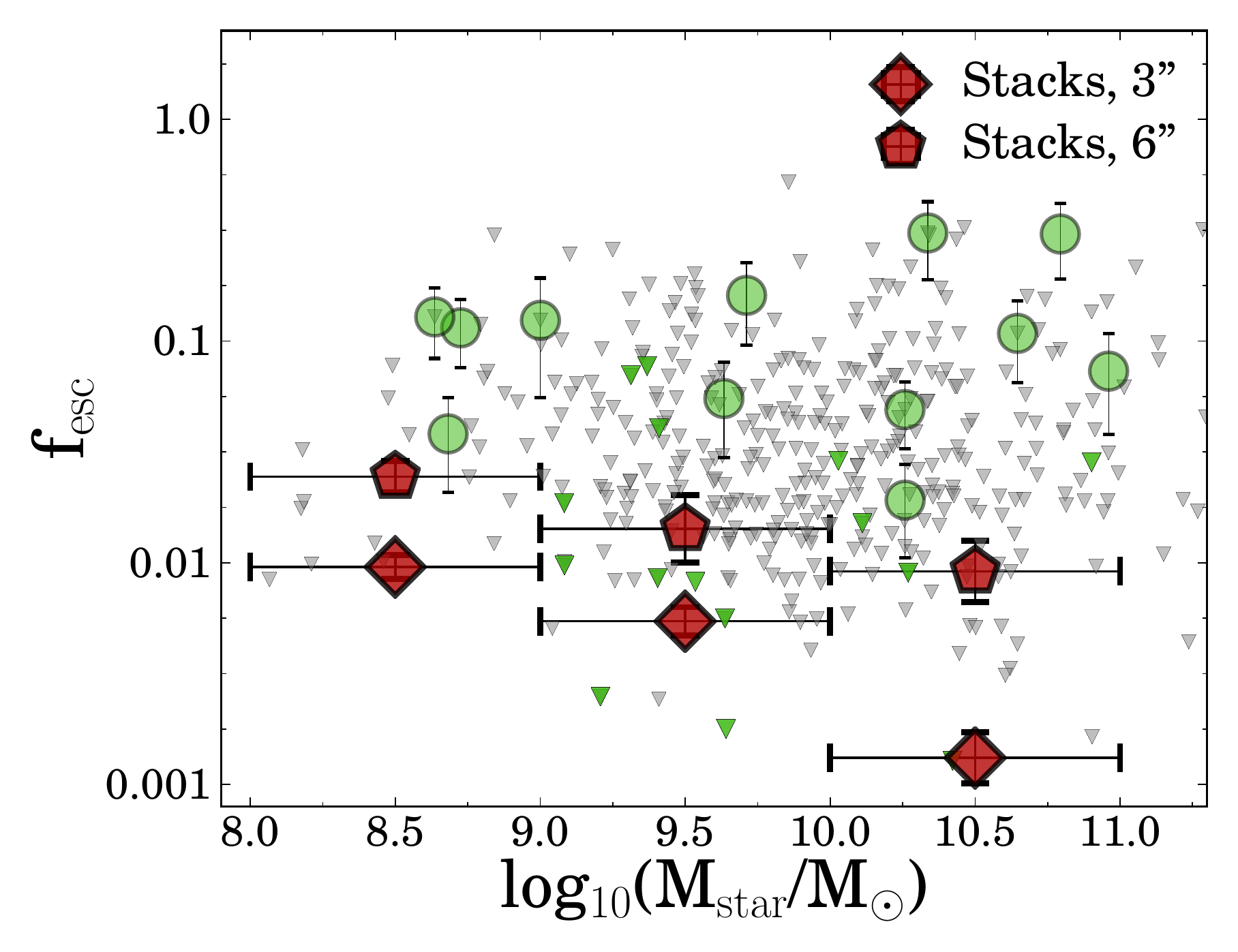}\\

\end{tabular}
    \caption{Ly$\alpha$ escape fraction versus SFR and stellar mass for galaxies without AGN for individual sources (left panels) and stacks (right panels). The green circles show our directly detected H$\alpha$-Ly$\alpha$ emitters, grey triangles highlight upper limits (green triangles have a UV detection in NB392) and our X-ray identified AGN with Ly$\alpha$ are shown in red diamonds. We also add the H$\alpha$ selected sample from \citet{Oteo2015} in orange stars and the Ly$\alpha$ selected sample from \citet{Song2014} as blue pentagons. Our survey clearly extends the probed parameter space in galaxy properties. Stacked values are shown for two different radial distances to the center (corresponding to 3$''$ diameter apertures - diamonds, and 6$''$ diameter apertures - pentagons). The typical measurement uncertainty in stellar mass is indicated in the {\it bottom left} panel. {\it Top row}: The {\it left} panel shows the SFR obtained from H$\alpha$ versus f$_{\rm esc}$ of individual sources. Although a correlation is expected by definition, it can be seen that on average, galaxies with a higher SFR have a lower escape fraction. The grey region in the {\it right} panel shows the SFRs typical for galaxies in the sample from \citealt{Hayes2010}, who inferred f$_{\rm esc} = 5.3\pm3.8$\%. {\it Bottom row:} Escape fraction versus stellar mass. While f$_{\rm esc}$ can be relatively high for low mass galaxies, the stacked results indicate that f$_{\rm esc}$ might decrease weakly with increasing stellar mass. The large difference in f$_{\rm esc}$ between some massive individual sources and the stacked values indicate that there is likely significant scatter in the values of f$_{\rm esc}$ at this mass range.} 
    \label{fig:correlations_sfr}
\end{figure*}

\subsection{{\sc[Oii]} and {\sc[Oiii]} emission lines}
In addition to NB$_K$ imaging, the HiZELS survey consists of NB$_J$ and NB$_H$ imaging in the same fields. These filters are designed such that they also cover the redshifted {\sc[Oii]} (in NB$_J$) and {\sc[Oiii]} (in NB$_H$) emission at $z=2.23$, similar to e.g. \cite{Nakajima2012} and \cite{Sobral2012}. Out of the 488 H$\alpha$ emitters, 23 and 70 galaxies are detected in {\sc[Oii]} and {\sc[Oiii]}, respectively. Two out of the 9 AGN are detected as line-emitters in both NBs, two as {\sc [Oiii]} emitters, and two AGN as {\sc [Oii]} emitters. The three remaining AGN all have a spectroscopic redshift at $z=2.21-2.23$ \citep{Brizels}. This means that all our AGN are either spectroscopically confirmed or have highly accurate photometric redshifts, with emission lines in at least two narrow-bands.

There are two star-forming HAEs with Ly$\alpha$ and {\sc[Oiii]} (ID 1993 and 2600, see Fig. $\ref{fig:dual_detections}$ and Table $\ref{tab:catalog}$). Compared to all HAEs with detection in {\sc[Oiii]} and similar limits on the Ly$\alpha$ flux and f$_{\rm esc}$, these galaxies have the lowest mass and SFR ($\sim 10^{8.6}$ M$_{\odot}$, SFR $< 8$M$_{\odot}$ yr$^{-1}$), most extreme UV slopes (either bluest, ID 2600, or reddest, ID 1993). They have f$_{\rm esc} \sim 10$\% and {\sc[Oiii]} EW$_0\sim 100$ {\AA}, resembling local Green pea galaxies \citep[e.g.][]{Henry2015}. Their dust corrections are uncertain, because the two methods described in \S 3.4.3 give a factor 3-5 difference depending on the method. Compared to the other Ly$\alpha$ detected HAEs, these galaxies have the highest Ly$\alpha$ EWs. ID 2600 has very high H$\alpha$ EW$_0$ of $\sim 960$ {\AA}, similar to the most extreme emission line galaxies \citep[e.g.][]{Amorin2015}, while the Ly$\alpha$ EW$_0$ is 80 {\AA}. ID 1993, however, has a relatively low H$\alpha$ EW$_0$ of 70 {\AA}, and Ly$\alpha$ EW$_0$ of 68 {\AA}. We note that if we use the \cite{GarnBest2010} dust correction, these two galaxies would have a factor 2-3 higher f$_{\rm esc}$. 

\section{Correlations between Ly$\alpha$ escape and galaxy properties}
\begin{figure*}
\begin{tabular}{cc}

\includegraphics[width=8.6cm]{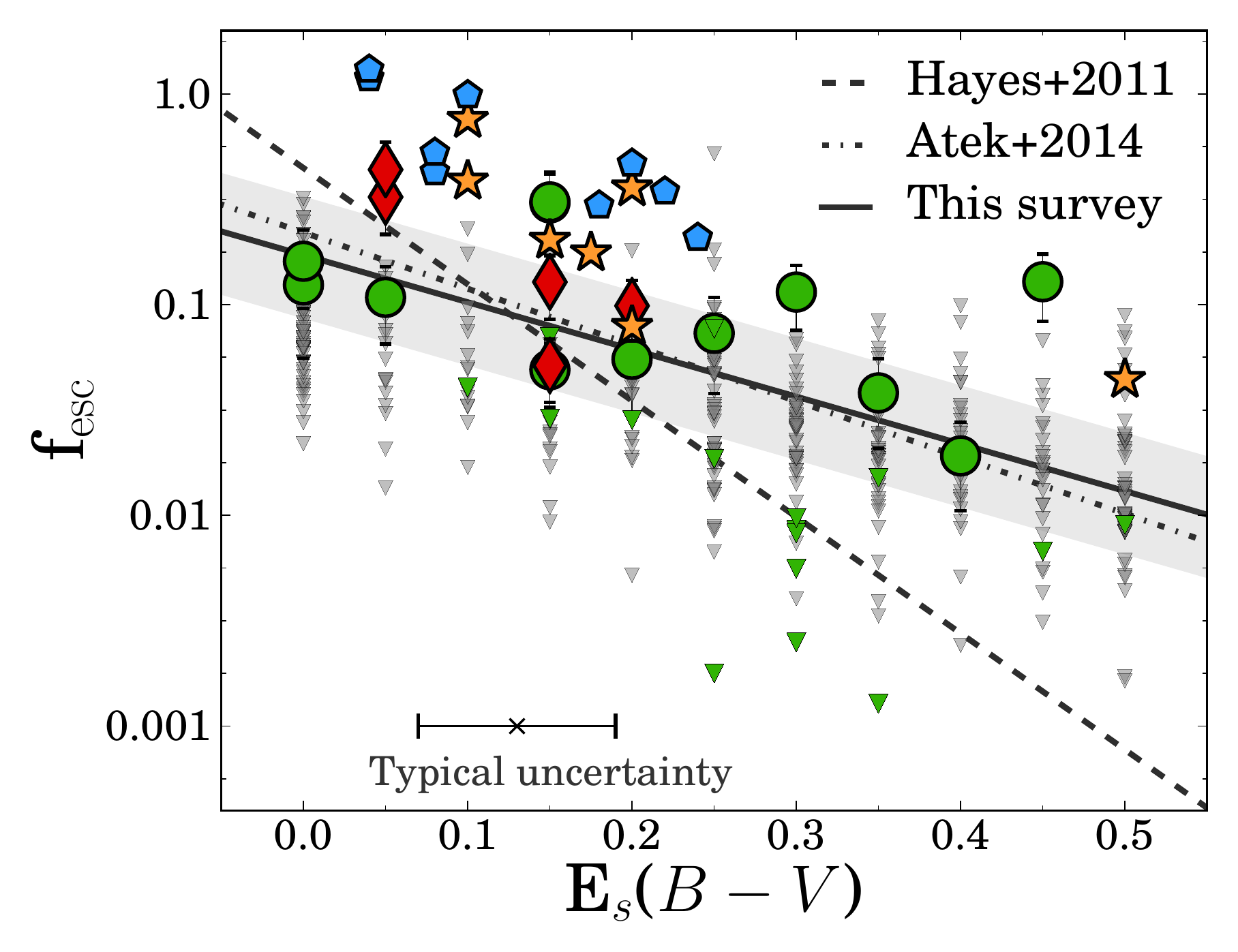}&

\includegraphics[width=8.6cm]{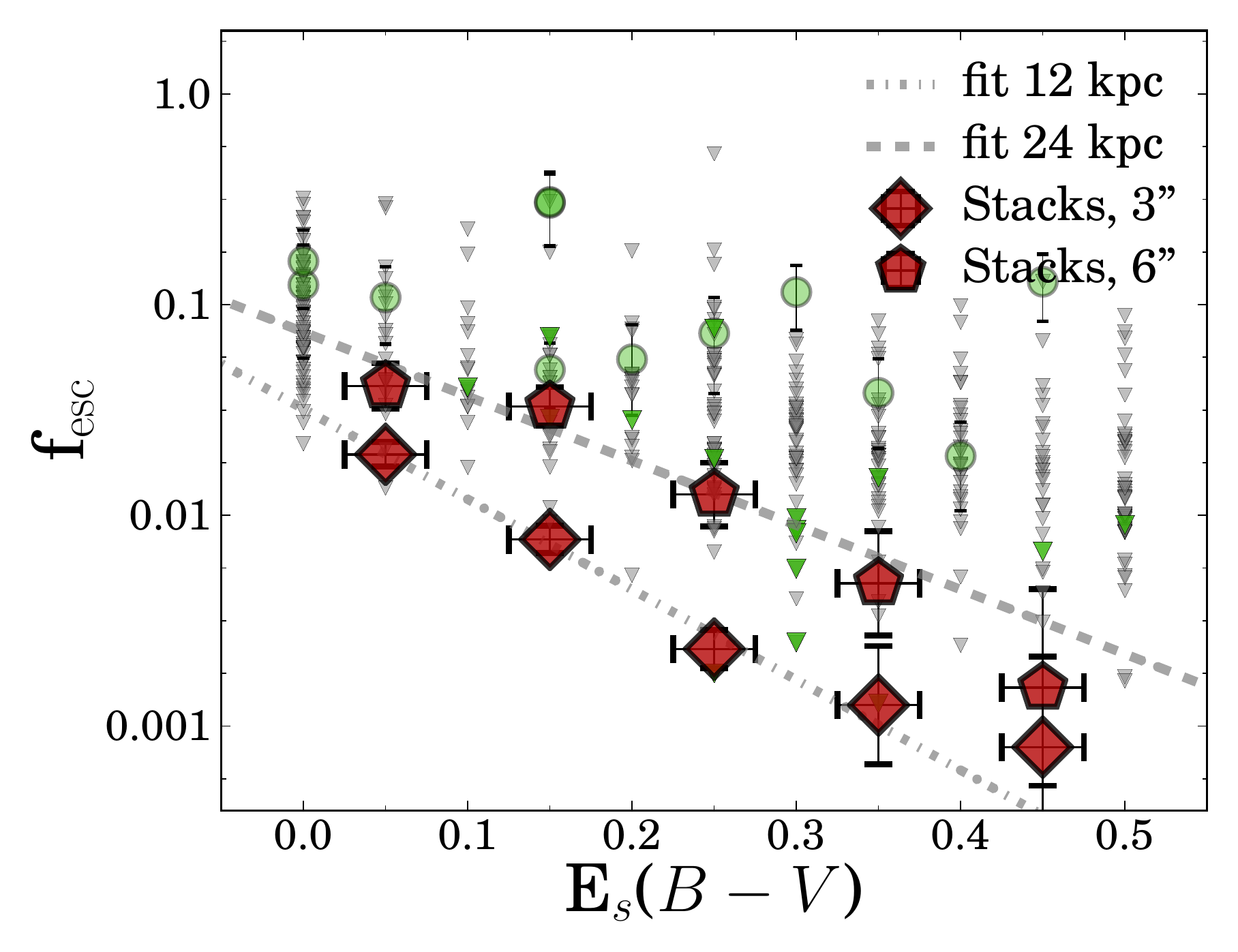}\\

\includegraphics[width=8.6cm]{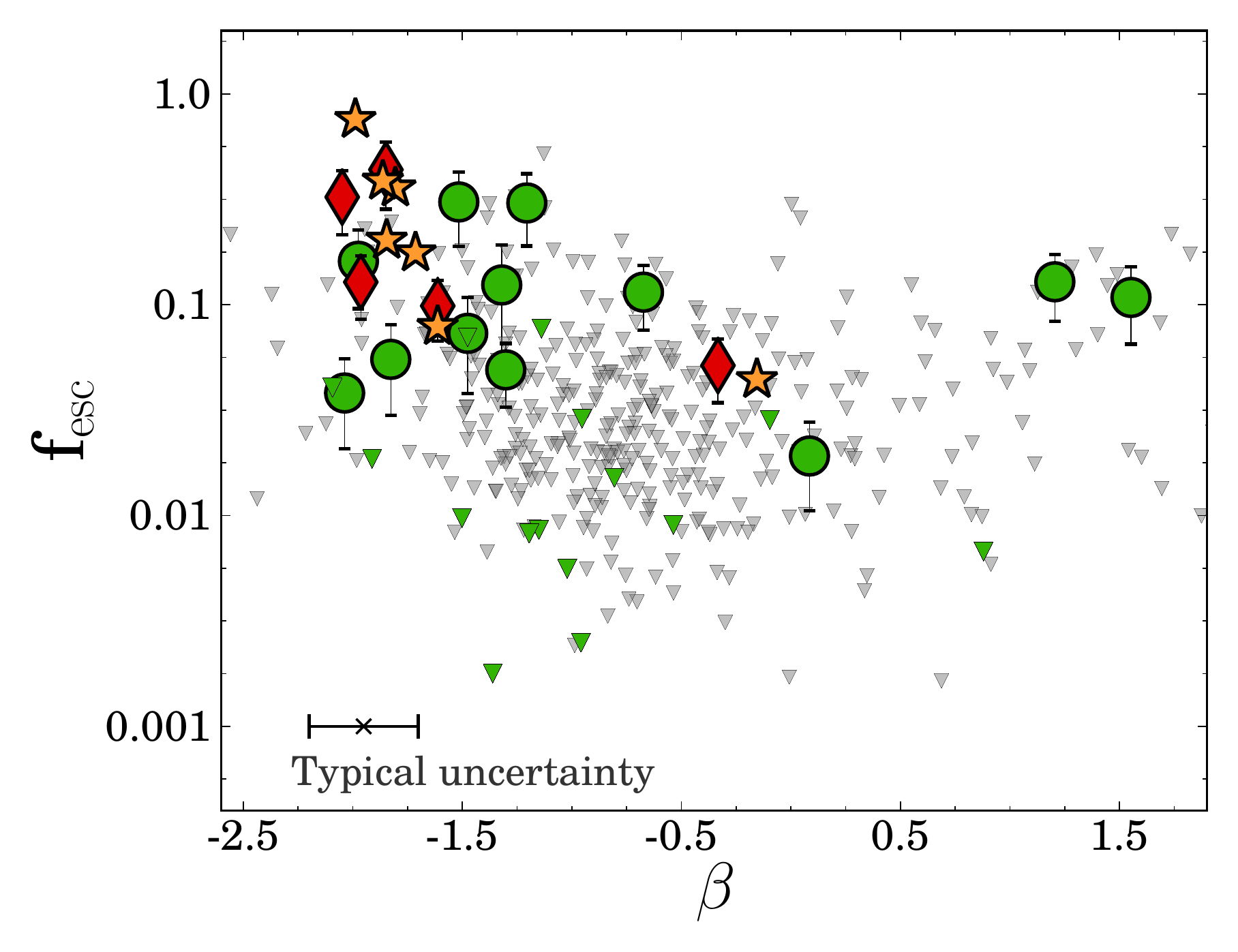}&

\includegraphics[width=8.6cm]{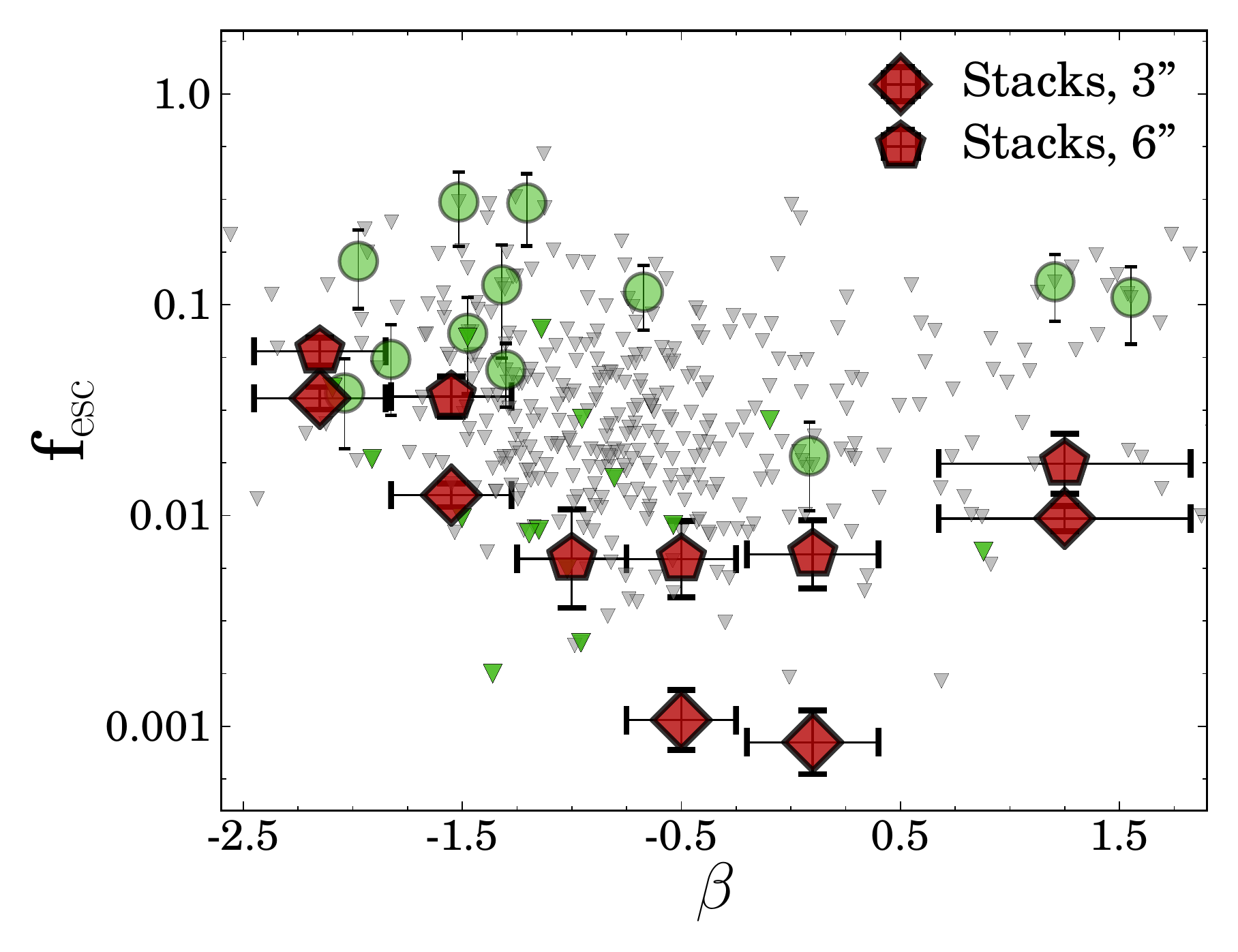}\\

\end{tabular}
    \caption{Correlations between f$_{\rm esc}$ and dust extinction, rest-frame $B-V$ colour and $\beta$, with symbols as defined in Fig. $\ref{fig:correlations_sfr}$. We indicate the maximum typical uncertainties on extinction and $\beta$ in the {\it left} panels. {\it Top row}: The {\it left} panel shows that f$_{\rm esc}$ is anti-correlated with the dust extinction, although there is significant scatter. This means that at fixed aperture, Ly$\alpha$ preferentially escapes galaxies with low dust content. For comparison, we show the relation at $z\sim2-3$ from \citet{Hayes2011} and $z\sim0.3$ from \citet{Atek2014}. Our best fitted relation (to detections and upper limits, see text) resembles more that of local galaxies. The grey band shows the 1$\sigma$ error of the normalisation of the fit. {\it Right:} Stacked values of escape fraction versus dust extinction. A similar trend is seen as for individual sources. However, because stacks are not biased towards high f$_{\rm esc}$ values, the normalisation is lower. {\it Bottom row:} Escape fraction versus $\beta$ for individual sources ({\it left}), which confirms that there is a bimodal relation between f$_{\rm esc}$ and galaxy UV colour, such that either very blue or very red galaxies emit significant Ly$\alpha$. As seen in stacks in the {\it right} panel, this bimodal trend is most clear at the largest apertures.}
    \label{fig:correlations_dust}
\end{figure*}

We use our sample of dual Ly$\alpha$-H$\alpha$ emitters and upper limits for the other 468 star-forming HAEs to search for potential correlations between galaxy properties and the Ly$\alpha$ escape fraction, shown in Fig. $\ref{fig:correlations_sfr}$ and Fig. $\ref{fig:correlations_dust}$. Our sample is compared with the H$\alpha$ selected sample of \cite{Oteo2015} and the spectroscopically, Ly$\alpha$-selected sample of \cite{Song2014}. The main difference with our sample is that it has deeper Ly$\alpha$ observations than \cite{Oteo2015} and spans a wider range of galaxy properties. The major difference in respect to \cite{Song2014} is that their sample is selected on strong Ly$\alpha$, and therefore biased towards sources with high f$_{\rm esc}$. 

In addition to studying correlations for individual sources, we also stack different subsets of galaxies. As described in \S 4, we will limit ourselves to the H$\alpha$ emitters with the deepest NB392 observations. We divide our sample by SFR(H$\alpha$), stellar mass, dust extinction and UV slope and ensure that our results do not depend on our particular choice of bin limits and width by perturbing both significantly. The benefit from studying correlations in bins of galaxies is that the results are less dependent on the systemic uncertainties in the dust corrections. While the systematic difference in dust corrections can be up to a factor of five for individual sources, the differences are much smaller over a statistical sample (compare for example Table $\ref{tab:numbers}$ and $\ref{tab:catalog}$). 

\subsection{Varying SFR(H$\alpha$) and Mass}
In the top row of Fig. $\ref{fig:correlations_sfr}$ we show that f$_{\rm esc}$ is anti-correlated with SFR. This is seen both in the individual sources and in stacks.  As both f$_{\rm esc}$ and SFR involve the H$\alpha$ measurement, we naively expected to find an anti-correlation between the two. Yet, the quantitative behaviour of this trend is not a priori trivial due to complex Ly$\alpha$ radiative transport. The qualitative trend is not strongly sensitive to the dust correction method applied. By fitting a linear relation to our stacked values (in logarithmic space), we obtain:
\begin{equation}
{\rm log}_{10}({\rm f}_{\rm esc}/(\%))= 1.34^{+0.12}_{-0.12} - 1.32^{+0.10}_{-0.10}\, {\rm log}_{10} ( \rm SFR/(M_{\odot}\rm yr^{-1})) 
\end{equation}
for a 12 kpc aperture radius, with $\chi^2_{\rm red} = 1.96$. At a radius of 24 kpc, the normalisation is $2.43^{+0.15}_{-0.15}$ and slope $-1.65^{+0.08}_{-0.09}$, with a $\chi^2_{\rm red} = 1.27$. This means that there is a $\sim13\sigma$ anti-correlation between the escape fraction and the SFR (as a slope of zero is rejected at 13$\sigma$). According to this relation, a galaxy with a SFR of 4 M$_{\odot}$ yr$^{-1}$ has a typical Ly$\alpha$ escape fraction of $\sim 3.5\pm1.8$\% (in 12 kpc apertures). We note that this relation probably turns over at lower SFR than we probe (as f$_{\rm esc}$ would otherwise reach values $>100$ \%). The top-right panel of Fig. $\ref{fig:correlations_sfr}$ also shows that the Ly$\alpha$ escape fraction is higher at larger radii at all SFRs, although the errors become a bit larger.  

It can be seen that our directly detected sample occupies the region in parameter space which has highest escape fraction at fixed SFR. Compared to the Ly$\alpha$ selected sample from \cite{Song2014}, we find a stronger anti-correlation between f$_{\rm esc}$ and SFR at a lower normalisation. This is because Ly$\alpha$ selected sources are biased towards high values of f$_{\rm esc}$. 

The bottom row of panels in Fig. $\ref{fig:correlations_sfr}$ shows how f$_{\rm esc}$ is related to stellar mass. While the stacked values show a weak anti-correlation (although at very low significance), there is no evidence for a trend between f$_{\rm esc}$ and stellar mass for the individually detected sources even though our sample clearly extends the probed dynamic range up to higher masses. As massive galaxies would naively be expected to have a lower f$_{\rm esc}$, since they tend to have a higher SFR and are dustier \citep[e.g.][]{Ibar2013}, this means that the Ly$\alpha$ escape fraction is not only determined by the scale of a galaxy and it is likely that more subtle processes such as dust and gas dynamics play an important role. Interestingly, the individual detected sources with high stellar mass are at much higher f$_{\rm esc}$ than the stacked values, indicating that there is significant scatter in f$_{\rm esc}$ in this mass range.

\subsection{Dust extinction and UV slope}
In the top row of Fig. $\ref{fig:correlations_dust}$ we show how f$_{\rm esc}$ is related to dust extinction. For individual sources and for stacks, we find that there is an anti-correlation between f$_{\rm esc}$ and dust extinction, although this relation contains significant scatter, caused in part by errors in our measurement of f$_{\rm esc}$ and E$(B-V)$. With our data, we are able to extend the previous found relation from \cite{Hayes2011} to lower escape fractions and higher dust extinctions. We fit the following linear relation:
\begin{equation}
{\rm f}_{\rm esc} = C \times 10^{-0.4 {\rm E}(B-V) k_{\rm Ly\alpha}}
\end{equation}
This fit is performed by simulating a large grid of normalisations and slopes and computing the $\chi^2$ for each combination of normalisation and slope in log(f$_{\rm esc}$) $-$ E$(B-V)$ space. Upper limits are taken into account by assigning them an f$_{\rm esc}$ of 0.01 \% and using their upper limit as error. Since a fit to individual galaxies is mostly determined by the directly detected dual H$\alpha$-Ly$\alpha$ emitters, which are biased towards high f$_{\rm esc}$, we also fit to the stacked values.  

By minimising the $\chi^2$ for individual galaxies, we find $C = 0.17^{+0.15}_{-0.09}$ and $k_{\rm Ly\alpha} =5.60^{+3.45}_{-2.90}$, such that a galaxy with E$(B-V) = 0$ has an escape fraction of 17\%. This is lower in normalisation (although the errors are significant), and significantly shallower in slope than the fit from \cite{Hayes2011}, who finds $C$ is 0.445 and $k_{\rm Ly\alpha} = 13.8$. The normalisation and slope are more similar to the $z=0.3$ result from \cite{Atek2014} ($C=0.22$, $k_{\rm Ly\alpha} = 6.67$). A possible explanation could be that \cite{Hayes2011} misses dusty galaxies, such that they infer a steeper slope, which would be consistent with the discussion in \cite{Oteo2015}. We discuss this further in \S 8.2.

\begin{figure*}
\begin{tabular}{cc}
\includegraphics[width=8.4cm]{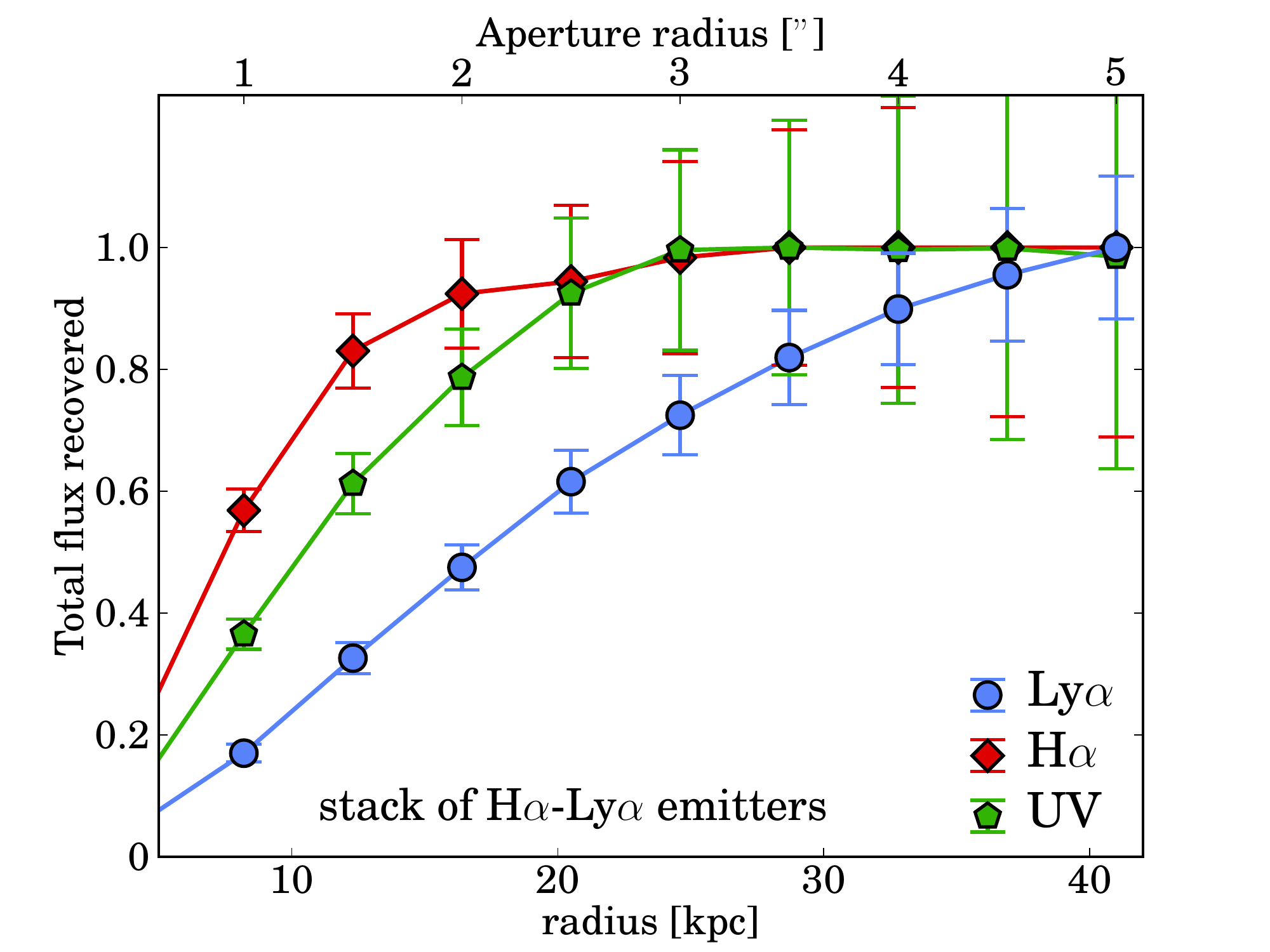}&
\includegraphics[width=8.4cm]{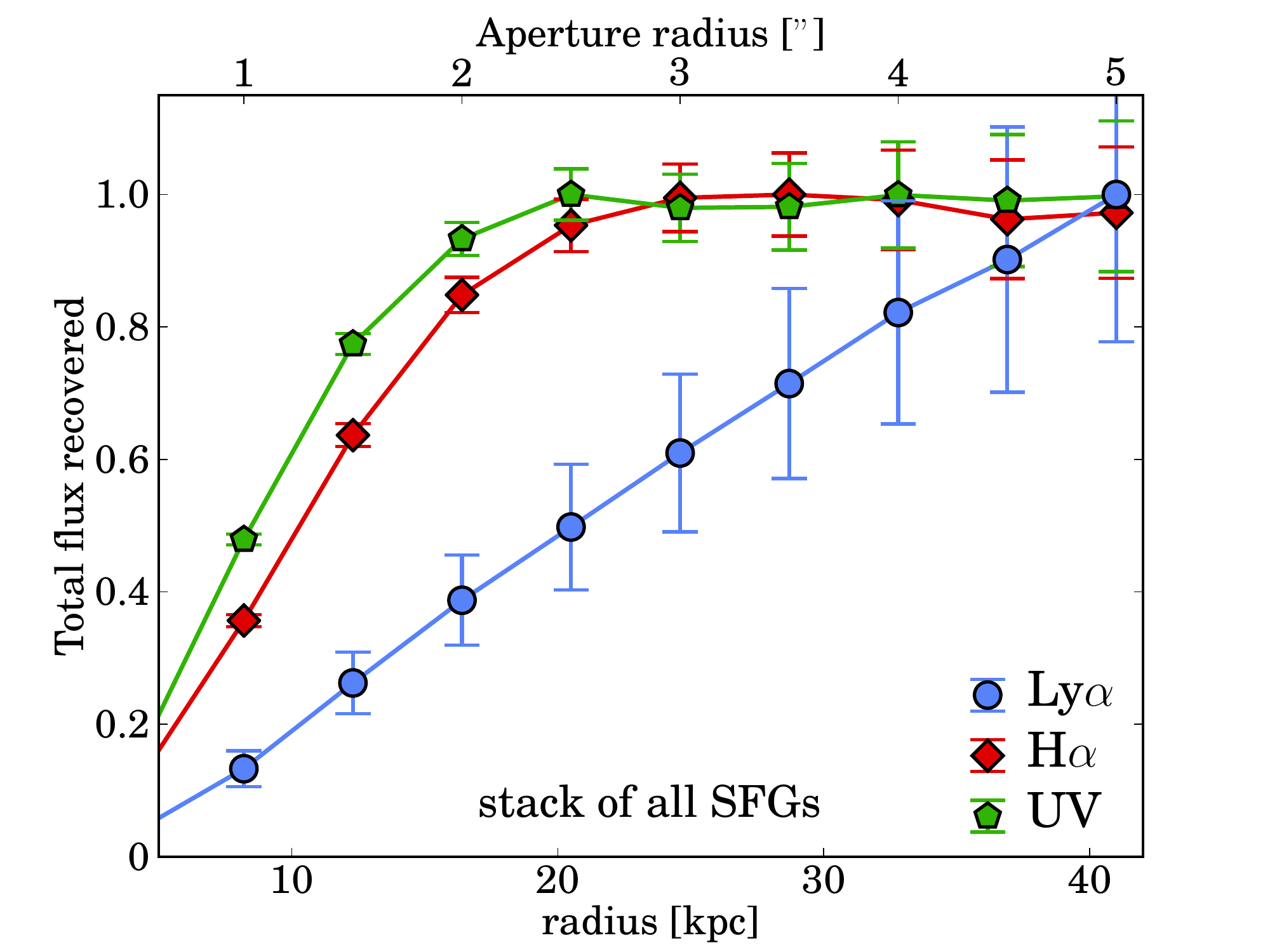}\\
\end{tabular}
\caption{Fraction of the total retrieved flux (defined as the maximum flux within 5$''$ radius) as a function of aperture for the UV, H$\alpha$ and Ly$\alpha$ and stack of direct detected H$\alpha$-Ly$\alpha$ emitters ({\it left}) and stack of all star-forming galaxies ({\it right}). It can be seen that while H$\alpha$ and the UV (from PSF convolved HST F814W imaging) quickly reaches the total flux in both cases, the Ly$\alpha$ flux continues to increase. For the direct detected sources, the Ly$\alpha$ flux increases somewhat more rapidly because sources where selected on bright central Ly$\alpha$. These sources also have more compact H$\alpha$ than UV emission. For the stack of all SFGs, Ly$\alpha$ continues to increase (although the errors are significant), such that deeper observations are required to test whether we have captured the full extent of Ly$\alpha$.}
\label{fig:growth_curves}
\end{figure*}
For stacks, we find $C = 0.03^{+0.01}_{-0.01}$ and $k_{\rm Ly\alpha} =10.71^{+0.89}_{-1.01}$ for 12 kpc apertures and $C = 0.08^{+0.02}_{-0.01}$ and $k_{\rm Ly\alpha} =7.64^{+1.38}_{-1.36}$ for 24 kpc apertures. Our fit to stacked data is less biased towards high values of f$_{\rm esc}$ and therefore at a lower normalisation. The slope is slightly higher, although still not as high as the slope inferred by \cite{Hayes2011}. Similar as seen for the stacks in bins of stellar mass, the individually detected galaxies at highest dust attenuations have much higher f$_{\rm esc}$ than the median stack. This means that there is a lot of scatter in the values of f$_{\rm esc}$ at the highest dust attenuations.

We furthermore note that part of the correlation between f$_{\rm esc}$ and E$(B-V)$ is expected because there is a dust correction in the H$\alpha$ flux, and thus in f$_{\rm esc}$. The fact that the correlation is rather weak (the slope is inconsistent with being zero at only 1.9$\sigma$ and as there is significant scatter), indicates that dust is not the only regulator of Ly$\alpha$ escape, a result already found by \cite{Atek2008} at low redshift. We also note that the trend between f$_{\rm esc}$ and E$(B-V)$ becomes somewhat bimodal when using the \cite{GarnBest2010} dust correction, meaning that there are galaxies with high dust attenuation and high f$_{\rm esc}$, which is virtually impossible with a \cite{Calzetti2000} dust correction. Galaxies with low E$(B-V)$ can in this case also have a lower f$_{\rm esc}$, leading to a flattening of the relation.

The bottom row of panels in Fig. $\ref{fig:correlations_dust}$ shows how f$_{\rm esc}$ is related with the UV slope $\beta$. As the UV slope is also a tracer of dust attenuation, we also find a tentative anti-correlation with escape fraction for galaxies with $\beta<0$, but for stacks and individual sources, particularly when the HAEs from \cite{Oteo2015} are included. Surprisingly, the trend between f$_{\rm esc}$ and $\beta$ seems to reverse for redder galaxies, leading to a bimodal relation which is particularly seen in measurements of f$_{\rm esc}$ with 3$''$ apertures. The (maximum) typical error of $\beta$ is indicated in the bottom left panel of Fig. $\ref{fig:correlations_dust}$ and is sufficiently small to exclude measurement errors of $\beta$ as the source of the observed bimodal behaviour. The bimodal behaviour is also seen when correcting for dust using the \cite{GarnBest2010} prescription.

\section{Extended emission}
\begin{figure*}
\begin{tabular}{cc}
	\includegraphics[width=8.2cm]{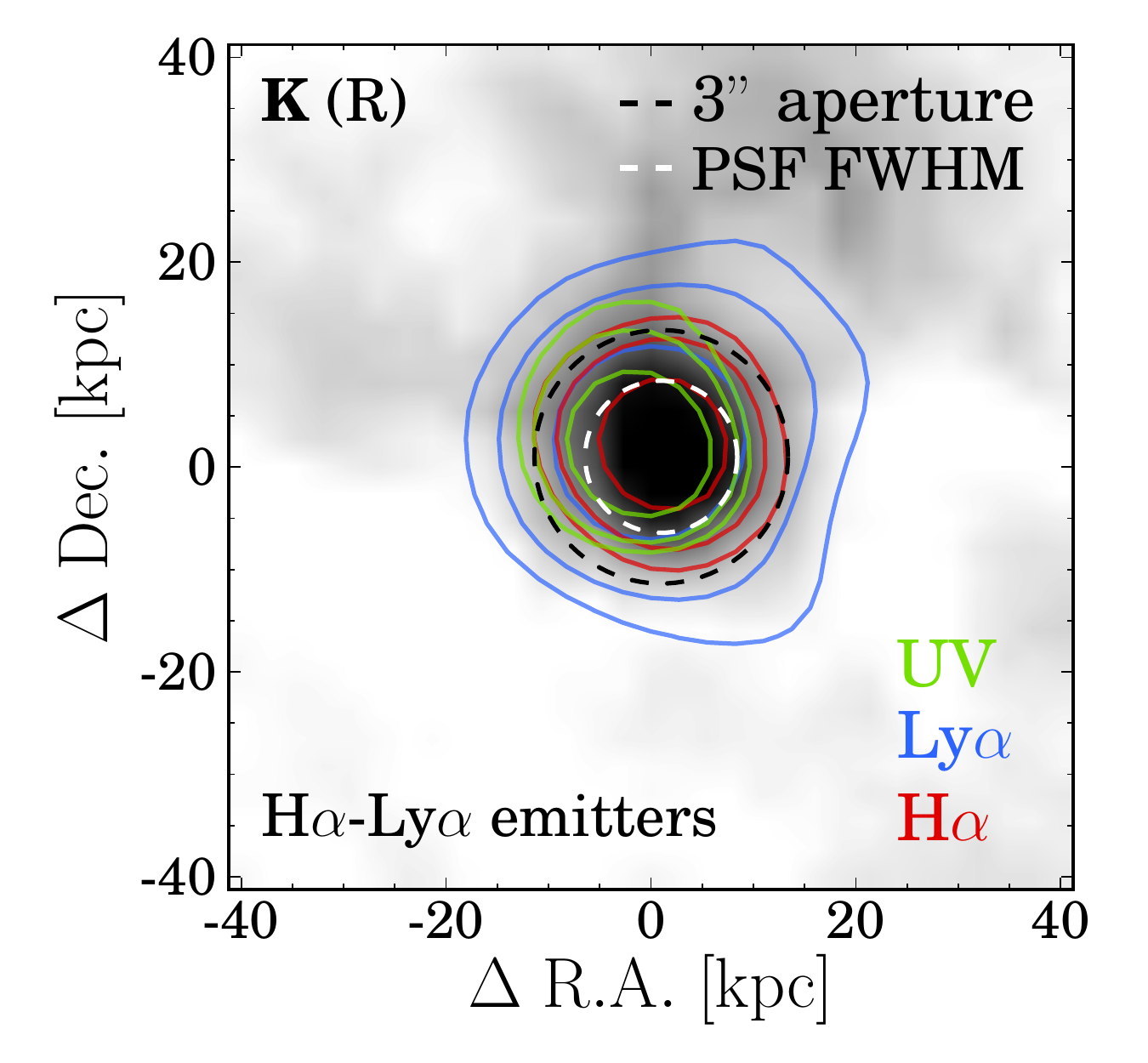} &
	\includegraphics[width=8.2cm]{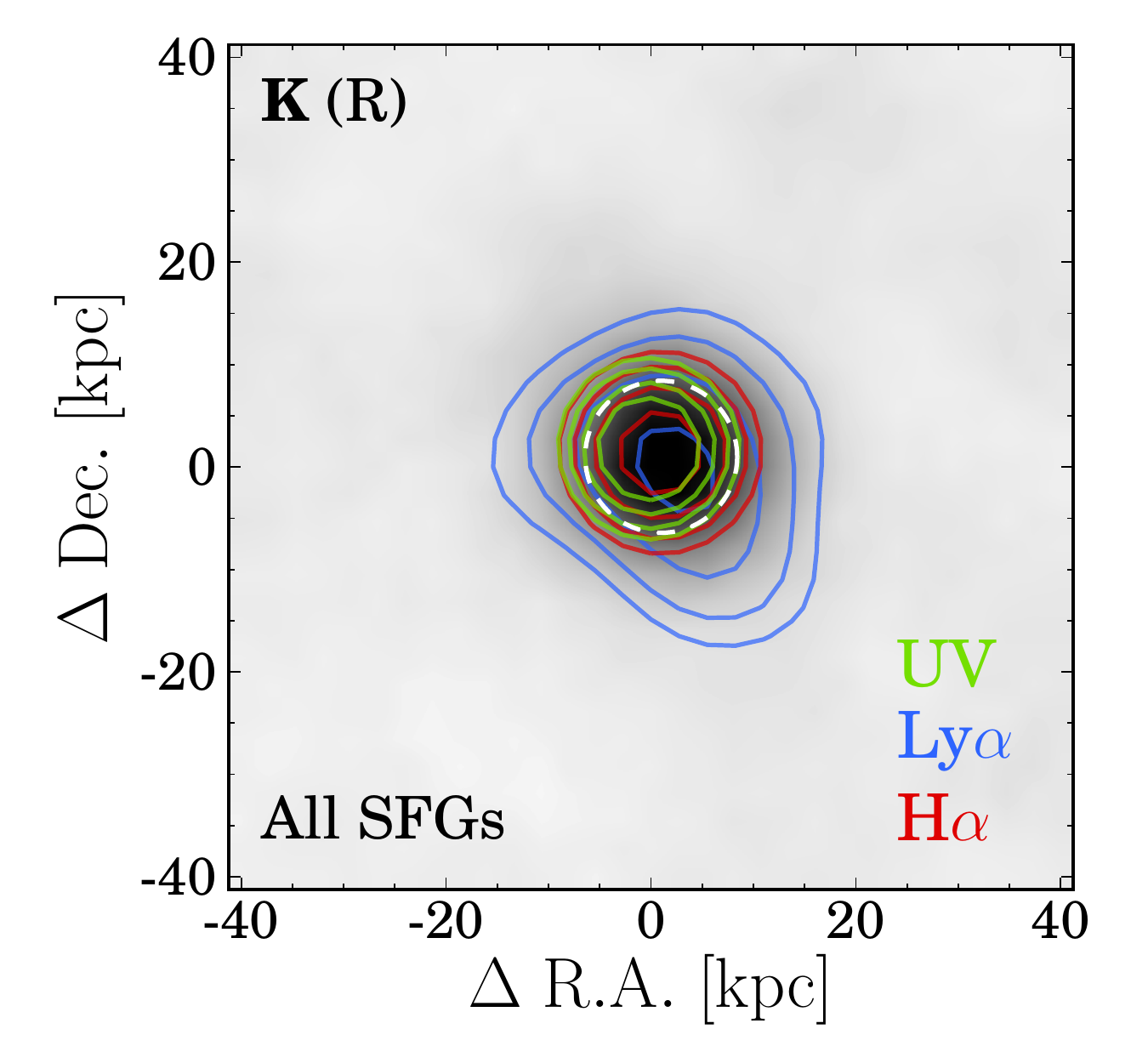} \\
\end{tabular}
    \caption{Thumbnails of the stacks of the 12 H$\alpha$-Ly$\alpha$ emitters (left) and full sample of SFGs (right). The background images shows the $K$ band (where we removed the contribution from H$\alpha$ using NB$_K$), which corresponds to rest-frame $R$. The Ly$\alpha$ emission is shown in blue contours, H$\alpha$ emission is shown in red and UV (from observed F814W, corresponding to $\sim 2000${\AA} rest-frame) in green. In the left panel we also indicate the 3$''$ diameter aperture used for measurements of individual sources in black dashed lines. The PSF FWHM is indicated with a white dashed circle in both panels. The contour levels are normalised to the peak flux in each band. The outer Ly$\alpha$ contour represents 7.5$\sigma$ for the {\it left} panel, and 3$\sigma$ for the {\it right} panel. This corresponds to 0.37 and 0.52 of the peak flux, respectively. The 1$\sigma$ surface brightness limit in the two panels is 9.0 and 2.5 $\times10^{-19}$ erg s$^{-1}$ cm$^{-2}$ arcsec$^{-2}$. Other contours correspond to a fraction of 0.5 and 0.75 of the peak flux in the {\it left} panel, and 0.6, 0.75 and 0.9 in the {\it right} panel. In both panels, it can be seen that H$\alpha$ traces the UV light very well. Ly$\alpha$ is more extended than H$\alpha$ and the UV for both the (biased) stack of direct detect H$\alpha$-Ly$\alpha$ and the full stack of SFGs, indicative of scattering. Ly$\alpha$ extends up to $\sim 20$ kpc distance from the center at the corresponding significances. The stack of the 12 H$\alpha$-Ly$\alpha$ emitters is extended up to 30 kpc at $3\sigma$ significance.}
    \label{fig:thumbnails}
\end{figure*}

As Ly$\alpha$ is a resonant emission line, scattering due to neutral hydrogen leads to a diffusion process similar to a random walk, which results in a lower surface brightness. Therefore, it is likely that in the presence of extended {\sc Hi}, Ly$\alpha$ emission will be more extended than the UV. Although our relatively large PSF FWHM and limited depth in NB392 ($\sim 1.8''$ and $\sim 24$ AB magnitude, respectively) limit the study of extended Ly$\alpha$ emission at low surface brightness, we can test how our measured escape fraction depends on the chosen aperture size. We do this by analysing the stacked images for the (biased) sample of 12 dual H$\alpha$-Ly$\alpha$ emitting SFGs, and for the stack of the 265 SFGs (see \S 5 and Table $\ref{tab:numbers}$). We measure both H$\alpha$, Ly$\alpha$ and the rest-frame UV in apertures ranging from $2-10''$ in diameter. 

As seen in Fig. $\ref{fig:growth_curves}$ (and illustrated by Fig. $\ref{fig:thumbnails}$), Ly$\alpha$ is significantly more extended than the UV (traced by convolved {\it HST} F814W imaging) and H$\alpha$ for both stacks. The stack of all SFGs (right panel of Fig. $\ref{fig:thumbnails}$) has extended Ly$\alpha$ emission up to $\sim 20$ kpc distance from the center, at 3$\sigma$. At this significance, the stack of the 12 dual H$\alpha$-Ly$\alpha$ emitters (left panel of Fig. $\ref{fig:thumbnails}$) is extended up to 30 kpc, and is clearly more extended than the aperture that we used for the sources individually.

The growth curves for H$\alpha$ and the UV are similar, quickly growing to the maximum at $\sim20$ kpc. Ly$\alpha$ however continues to increase. The Ly$\alpha$ flux for the stack of the sample of 12 H$\alpha$-Ly$\alpha$ emitters peaks at $\sim$4$''$ radial apertures, as further increase in the fraction of recovered flux with increasing aperture is within the errors. The f$_{\rm esc}$ within this radius, $\sim30$ kpc, is $14.2\pm1.9$\%. At a radius of 1.5$''$ (similar to the aperture used for individual sources), we measure a stacked f$_{\rm esc}$ of $7.7\pm0.9$\%. At this aperture, only $\sim50$\% of the maximum observed Ly$\alpha$ flux is retrieved. The Ly$\alpha$ flux for the stack of all SFGs also continues to increase up to at least 30 kpc. By fitting a linear relation between the fraction of the total recovered flux and radius, we find that r$_{90}$ (the radius at which 90 \% of the flux is retrieved) for our stack of directly detected H$\alpha$-Ly$\alpha$ is 31.3$\pm$1.5 kpc. For the stack of all SFGs, r$_{90} = 36.0\pm3.8$ kpc. For H$\alpha$, we find values of r$_{90}=19.3\pm1.6$ kpc and r$_{90}=21.0\pm0.5$ kpc, respectively.

\begin{figure*}
\begin{tabular}{cc}
\includegraphics[width=8.3cm]{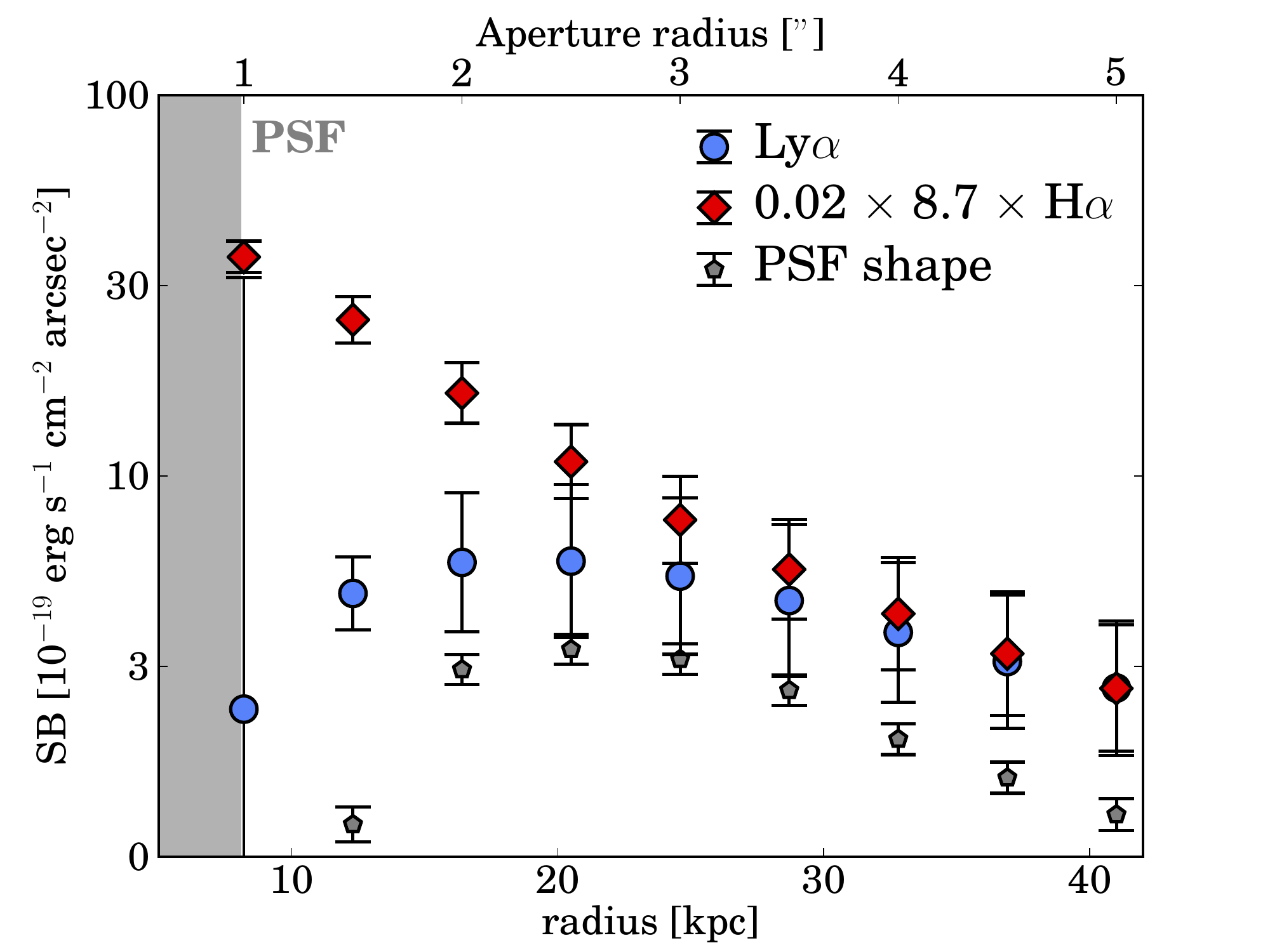}&
\includegraphics[width=8.3cm]{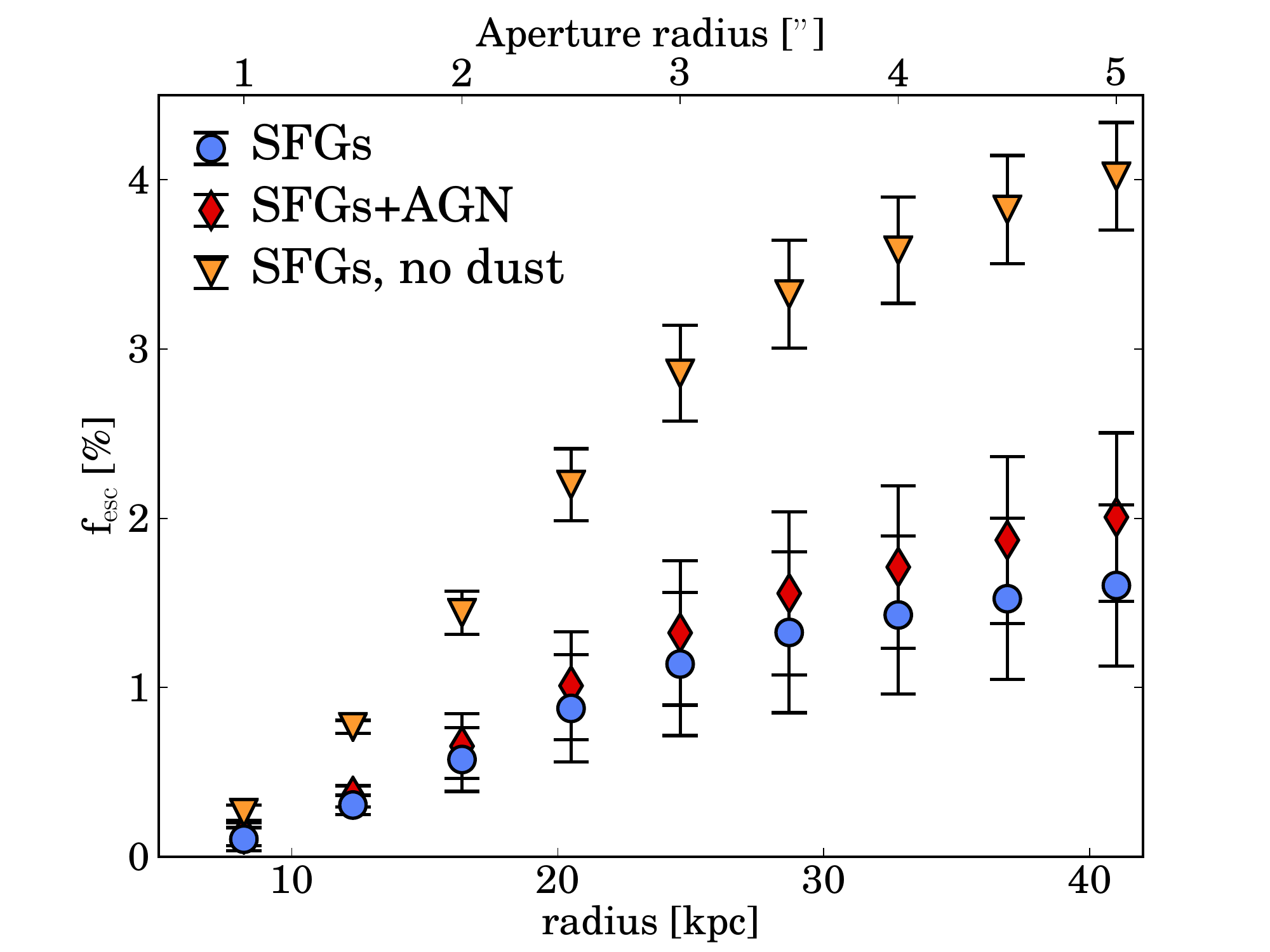}\\
\end{tabular}
\caption{{\it Left}: Surface brightness of our full stack of star-forming galaxies for H$\alpha$ and Ly$\alpha$ and we also show the surface brightness for the stack of the PSF control sample (\S 4.1), for which the stack of SFGs is corrected. The surface brightness that we observe is attributed to different shapes of the PSF of the NB and broadband. This surface brightness difference is added to the error of the Ly$\alpha$ surface brightness. We scaled the H$\alpha$ flux such that it has the same surface brightness as Ly$\alpha$ at a radius of 5$''$, corresponding to 41 kpc. Note that this scaling corresponds to an escape fraction of 2 \%, discussed in \S 7.1. {\it Right}: Escape fraction of the stack of all sources as a function of radial distance from the center. We compare stacking sources with and without removal of AGN, and we also show the escape fraction if we do not correct H$\alpha$ for dust. The escape fraction at large radii is slightly higher without the removal of AGN, because the AGN are typically bright in Ly$\alpha$ and the haloes around them have a larger scale.}
\label{fig:aperture_variations}
\end{figure*}

We show the surface brightness (SB) profile of the full stack of SFGs in the left panel of Fig. $\ref{fig:aperture_variations}$, where we scaled H$\alpha$ such that it has a similar SB as Ly$\alpha$ at 5$''$ radius, corresponding to an escape fraction of 2\% (see also the right panel). The H$\alpha$ profile follows an exponential (as the y-axis is logarithmic), decreasing with increasing radius. While the Ly$\alpha$ profile seems to be more complex, we note that the errors due to the different PSF shapes of broadband and NB are important, particularly in the centre part. Towards higher radii ($>30$ kpc), the Ly$\alpha$ signal from the stack is significantly above the errors due to differences in the PSF, and is thus real. We find no evidence that the integrated Ly$\alpha$ flux does not continue to grow up to 40 kpc distance, indicating that it can be extended up to larger radii if deeper surface brightness limits are reached. This also means that we can not yet directly infer the total f$_{\rm esc}$. 

Comparing the H$\alpha$ and Ly$\alpha$ profiles results in an increasing Ly$\alpha$ escape fraction with increasing aperture (see the right panel of Fig. $\ref{fig:aperture_variations}$). The f$_{\rm esc}$ increases from $0.3\pm0.05$\% at 12 kpc distance to $1.6\pm0.5$\% at $\sim30$ kpc. Note that without dust correction f$_{\rm esc}$ is roughly a factor 2 higher. At the radius of 30 kpc, the Ly$\alpha$ surface brightness is $\sim 6\times10^{-19}$ erg s$^{-1}$ cm$^{-2}$ arcsec$^{-2}$ (see also the left panel), such that the extended emission is detected at $\sim2.4\sigma$ (at 2$\sigma$ confidence level, extended emission is seen up to $\sim 40$ kpc). As seen in the right panel of Fig. $\ref{fig:growth_curves}$ and illustrated in the right panel of Fig. $\ref{fig:thumbnails}$, Ly$\alpha$ is extended up to $\sim20$ kpc at $3\sigma$ confidence level. At 2$\sigma$ significance, it is extended up to $\sim30$ kpc. This means that aperture based measurements (including slit spectroscopy) might miss parts of Ly$\alpha$ emission, and that IFU's or specially designed NB filters are more suited.

\section{Discussion}
\subsection{A consensus on the value of f$_{\rm esc}$}
To date, a number of papers have been published on measuring the Ly$\alpha$ escape fraction with different selections of galaxies and methods. Typically, Ly$\alpha$ selected galaxies at $z>2$ have resulted in high escape fractions of $\sim 30$\% \citep[e.g.][]{Blanc2011,Nakajima2012,Wardlow2014,Kusakabe2014,Trainor2015}, even though techniques to estimate the intrinsic Ly$\alpha$ production range from UV to dual NB to FIR and spectroscopy. However, a Ly$\alpha$ selected sample of galaxies is not representative for the full (star-forming) galaxy population and estimates of f$_{\rm esc}$ are biased towards high values (as for example seen in Fig. $\ref{fig:correlations_sfr}$).  

With UV or rest-frame optical emission line selected galaxies, the typical f$_{\rm esc}$ at $z=2-3$ is around f$_{\rm esc}\sim$ 3-5\%, \citep[e.g.][]{Hayes2010,Kornei2010,Ciardullo2014}, but f$_{\rm esc}$ is found to increase with increasing redshift, up to $\sim 30$\% at $z=5.7$ \citep[e.g.][]{Hayes2011}. Our measured median value of 1.6$\pm0.5$\% of HAEs is lower than the value in these papers in the literature. However, we note that our measurement is the first which is independent of assumptions on the shape of the luminosity function and integration limits. Furthermore, as we will show now, the results are fully consistent with literature results when we account for different selections of galaxies and the different parameter spaces probed by different surveys.

Most samples have either been UV selected or selected with emission lines bluer than H$\alpha$ (e.g. {\sc [Oiii]}; \citealt{Ciardullo2014}), such that they might miss a population of dusty galaxies. This might cause a bias towards high f$_{\rm esc}$ values, as we show here that dustier galaxies typically have a lower f$_{\rm esc}$ (see Fig. $\ref{fig:correlations_dust}$, but note that we also observe significant f$_{\rm esc}$ for some dusty galaxies). On the contrary, \cite{Cassata2015} report a low fraction of UV selected galaxies with Ly$\alpha$ emission at $2<z<6$, corresponding to f$_{\rm esc} < 1$\%, concluding that the bulk of the Ly$\alpha$ luminosity density is coming from fainter galaxies than the typically UV bright galaxies that were targeted. This is consistent with our results, as these galaxies have a typical SFR of $\sim50-100$ M$_{\odot}$ yr$^{-1}$ \citep{Tasca2015}, and we show that the typical escape fraction at these SFRs is very low ($<0.5$\%).

The major difference between our survey and the matched H$\alpha$-Ly$\alpha$ survey at $z=2.2$ from \cite{Hayes2010} is that our HAEs probe a larger range in SFRs and stellar masses, because our observations probe a much larger volume ($\sim80$ times larger, yet with shallower depth). We have shown in the top row of panel of Fig. $\ref{fig:correlations_sfr}$ that the SFR anti-correlates with the escape fraction. If we compute SFRs from the H$\alpha$ emitters from \cite{Hayes2010} as described in \S 3.5, we find that their SFRs are between 0.4-4.4 M$_{\odot}$ yr$^{-1}$, illustrated as the grey region in the top-right panel of Fig. $\ref{fig:correlations_sfr}$. The difference between the inferred escape fraction of 5.3$\pm3.8$\% by \cite{Hayes2010} and our stack of SFGs can be fully explained by the different parts of parameter space covered, and the anti-correlation between f$_{\rm esc}$ and SFR. This also means that the bulk of Ly$\alpha$ luminosity density at $z=2.23$ is not emitted by star-forming galaxies of $>3$ M$_{\odot}$ yr$^{-1}$ (roughly the lower limit of the H$\alpha$ survey), but by much fainter galaxies. We note that \cite{Konno2015} infer a global f$_{\rm esc}$ of $\sim1.5$\% based on more recent measurements of UV and Ly$\alpha$ luminosity functions. 

\subsection{Dependence on galaxy properties}
In \S 6 we have shown that f$_{\rm esc}$ generally increases with decreasing SFR and dust attenuation (albeit with significant scatter), and that it shows bimodal behaviour with UV slope. The trends between f$_{\rm esc}$ and SFR and dust attenuation are also seen in the local Universe \citep{Hayes2014}. In this subsection we discuss how our inferred trends compare to previous results and speculate on the origin of bimodal relations.

\subsubsection{Relation with dust attenuation}
Other surveys have also established a correlation between f$_{\rm esc}$ and dust attenuation, both in the local Universe \citep{Atek2008,Atek2014} and at high redshift \citep{Verhamme2008,Kornei2010,Hayes2010,Blanc2011,Cassata2015}. This trend is in line with the significantly lower observed f$_{\rm esc}$ in the low redshift Universe \citep[e.g. $\sim 1$\% at $z=0.3-1$][]{Deharveng2008,Cowie2010,Wold2014}, as the dust content of galaxies increases with cosmic time \citep[c.f.][]{Hayes2011}. 

We find a weak trend between f$_{\rm esc}$ and dust attenuation, both for stacks and individual galaxies. As shown in the top row of panels of Fig. $\ref{fig:correlations_dust}$, our trends between f$_{\rm esc}$ and E$(B-V)$ are significantly shallower than the trend inferred by \cite{Hayes2011}, and resembles the trend seen in the local Universe \citep{Atek2014}. However, note that comparison is limited by systematic uncertainties and differences between the methods used to estimate f$_{\rm esc}$ (for example, \citealt{Atek2014} uses Balmer decrements to estimate the dust uncertainty, while our survey is dependent on SED fitting). While \cite{Hayes2010} only included H$\alpha$ and Ly$\alpha$ selected galaxies in their analysis, \cite{Hayes2011} added the UV selected galaxies at $z\sim3$ from \cite{Kornei2010} in order to have a larger sample. However, the UV selected sample requires additional uncertainties on the intrinsic production of Ly$\alpha$ and is biased towards lower dust attenuations (even though their mass and SFR distribution is similar to our H$\alpha$ selected sample).

We speculate that the prime origin of the discrepancy between our fitted trend and the trends from \cite{Hayes2010} and \cite{Hayes2011} is the combination that we probe more luminous galaxies with higher masses and SFRs (compared to their H$\alpha$ selected sub-sample) and more dusty galaxies (compared to their UV selected sub-sample), and therefore find a lower normalisation. As our parameter space also includes more dusty galaxies with significant f$_{\rm esc}$, the relation is flattened. We also note that we find that at a fixed dust attenuation, galaxies with a relatively higher escape fraction have a higher SFR, and are thus also more likely to not be present in the \cite{Hayes2011} sample (as galaxies with higher SFR are rarer), and leading to a flattening of the relation. At the highest dust attenuations, we observe individual galaxies with f$_{\rm esc}$ over two orders of magnitude higher than the typical value for galaxies with similar dust attenuation, indicating that there is a lot of dispersion in the values of f$_{\rm esc}$. 

It is interesting to note that we find that the E$(B-V)$ values of HAEs are not dependent on Ly$\alpha$ EW. The relation we would infer between f$_{\rm esc}$ and E$(B-V)$ for HAEs with Ly$\alpha$ EW$_0>25${\AA} would be flatter than the relation inferred for our full sample of HAEs. This is because dusty HAEs with high f$_{\rm esc}$ tend to have high Ly$\alpha$ EW$_0$, which may be a sign of outflows. Note however that we do not include Ly$\alpha$ selected sources with faint H$\alpha$ (and thus high f$_{\rm esc}$) in this analysis.

There still exist large systematic uncertainties on the method to correct for dust attenuation at high redshift. For example, if the nebular extinction is indeed stronger than the stellar extinction at high SFRs \citep[e.g.][]{Reddy2015}, this may influence our observed trends in Figures $\ref{fig:correlations_sfr}$ and $\ref{fig:correlations_dust}$. We use the \cite{GarnBest2010} prescription for dust attenuation to evaluate how such a differential dust correction affects our results, as this prescription is based on stellar mass and thus qualitatively similar to a dust attenuation which varies with SFR. On a source-by-source basis, the f$_{\rm esc}$ can vary significantly (see for example the difference in attenuations in Table $\ref{tab:catalog}$), however, the results for most stacks are very similar. The largest changes are seen for the stacks with either the highest or lowest SFRs, stellar masses and E$(B-V)$ values. We find that the relation between f$_{\rm esc}$ and E$(B-V)_{\rm star}$ flattens somewhat, which is driven by a higher attenuation (and thus lower f$_{\rm esc}$) for sources in the bin with lowest E$(B-V)$ values. The relations between f$_{\rm esc}$ and SFR and stellar mass would be steeper. Regardless of the method used to correct for dust, the trend we find between f$_{\rm esc}$ and E$(B-V)$ is relatively weak and there is a large dispersion in the values of f$_{\rm esc}$. This indicates that dust attenuation alone is not the most important regulator of f$_{\rm esc}$.

\subsubsection{Absence of relations and bimodality?}
Additional evidence that dust is not the most important (or at least not the only) regulator of f$_{\rm esc}$ is that there exists a significant population of red, dusty Ly$\alpha$ emitters (e.g. \citealt{Blanc2011,Guaita2011,Nilsson2011,Oteo2015,Taniguchi2015} and this survey), particularly in the presence of strong outflows \citep[e.g.][]{Atek2008,Martin2015}. We find tentative evidence that there are bimodal trends between f$_{\rm esc}$ and UV slope. This is supported by the observation that the two reddest galaxies in the sample (IDs 1057 and 1993) have a high f$_{\rm esc}$. These sources are surprisingly different: while 1993 has low mass and high dust content, 1057 is massive with little dust. These two sources are not atypical, in fact, almost all Ly$\alpha$ detected HAEs are either bluer or redder than the average HAE without Ly$\alpha$ detection (which have $\beta\sim-1$). Apart from having high Ly$\alpha$ EW, we find no galaxy property which is related to having a red UV slope and high escape fraction. We note that this bimodal trend can only be seen for samples which include red, relatively massive objects. The trend is therefore still consistent with the results from \cite{Hagen2015}, who find no statistical difference in the UV slopes of Ly$\alpha$ or {\sc [Oiii]} selected galaxies.

Fig. $\ref{fig:correlations_sfr}$ shows that there is only a weak trend between f$_{\rm esc}$ and mass for stacks, such that high mass galaxies have lower f$_{\rm esc}$. However, there are relatively massive galaxies which are detected with high f$_{\rm esc}$. This means that there is a lot of dispersion in the values of f$_{\rm esc}$ at high stellar masses. This is also seen in the galaxies with highest dust attenuation. It is interesting to note that the mass scale at which we find higher f$_{\rm esc}$ for individual sources coincides with the point where our H$\alpha$ selection starts picking out galaxies below the main relation between stellar mass and SFR, see also Fig. $\ref{fig:sfr_mstar}$. Because of this, and with observations of local galaxies with strong outflows and Ly$\alpha$ emission in mind, we speculate that this enhanced f$_{\rm esc}$ might be related due to outflows of (dusty) gas. Outflows in turn redshift Ly$\alpha$ out of resonance wavelengths, which facilitate the escape of Ly$\alpha$ photons, as for example observed in local galaxies \citep[e.g.][]{Bik2015,Duval2015,Herenz2015,RiveraThorsen2015}. This idea can be tested with spectroscopic follow-up.

\subsection{Extended Ly$\alpha$ emission}
There is increasing evidence that star-forming galaxies are surrounded by a low surface brightness Ly$\alpha$ halo, as indicated from stacks of UV selected galaxies \citep{Steidel2011} and Ly$\alpha$ emitters \citep{Matsuda2012,Feldmeier2013,Momose2014}. Ly$\alpha$ haloes have also been detected around faint individual Ly$\alpha$ selected galaxies, both locally and at high redshift \citep{Hayes2013,Wisotzki2015,Patricio2015,Caminha2015}. Moreover, luminous Ly$\alpha$ emitters, without clear signs of AGN activity, can also be extended up to at least $\sim 16$ kpc \citep[e.g.][]{Hayashino2004,Ouchi2013,Sobral2015CR7}. 

 As Ly$\alpha$ emission is observed to be more extended than the UV, it is usually attributed to resonant scattering \citep[e.g.][]{Zheng2010,Dijkstra2012,Hayes2013}, although simulations from \cite{Lake2015} indicate that collisional ionisation due to cooling flows might also contribute. Very extended Ly$\alpha$ emission on scales of $\sim100$ kpc has been observed in Ly$\alpha$ blobs \citep[e.g.][]{Matsuda2004} and Ly$\alpha$ haloes around radio galaxies \citep[e.g.][]{Rottgering1995,Venemans2007,Swinbank2015}, believed to be powered by either star formation or AGN activity \citep[e.g.][]{Geach2009,Overzier2013,Ao2015,Umehata2015}.

We find that the average H$\alpha$ selected, star-forming galaxy at $z=2.23$ shows Ly$\alpha$ emission which is more extended than their H$\alpha$ or the UV emission (see Fig. $\ref{fig:growth_curves}$, Fig. $\ref{fig:thumbnails}$ and Fig. $\ref{fig:aperture_variations}$). These galaxies are typically dustier than the sources from \cite{Steidel2011} (with our sample of HAEs having a median A$_{\rm Ly\alpha} = 3.0$, compared to A$_{\rm Ly\alpha} = 1.12$ from \citealt{Steidel2011}). The surface brightness of the stack of all sources is therefore fainter (roughly a factor two, at a radial distance of 20 kpc) than the stack by \cite{Steidel2011}, but we note that the stack of galaxies with E$(B-V) \sim 0.1$ (A$_{\rm Ly\alpha} = 1.2$) has similar Ly$\alpha$ surface brightness values as the stack from \cite{Steidel2011}. We also note that our stack of SFGs is fainter in the UV by 0.5 magnitude, but has a dust corrected SFR of $\sim42$ M$_{\odot}$yr$^{-1}$, compared to the typical SFR of $\sim34$ M$_{\odot}$yr$^{-1}$ of the sample from \cite{Steidel2011}, because of the higher dust attenuation of our sample. 

With the current depth, we can not measure the maximum extent of Ly$\alpha$ for the stack of all SFGs or the stack of directly detected H$\alpha$-Ly$\alpha$ emitters (see Fig. $\ref{fig:growth_curves}$). By comparing the ratio of r$_{90}$ of Ly$\alpha$ and H$\alpha$ for both stacks, we find that Ly$\alpha$ radii are larger than H$\alpha$ radii with a factor of $1.6-1.7$. These values are comparable to those found for local Ly$\alpha$ analogs from \cite{Hayes2013}, although that analysis used r$_{20}$, which is undesirable for our sample due to the limitations from the PSF. For these local analogs, the average ratio between the Ly$\alpha$ and H$\alpha$ radii is however slightly higher than the ratio for our stack of SFGs, which may indicate that we have not yet observed the full extent of Ly$\alpha$.

For the stack of directly detected H$\alpha$-Ly$\alpha$ emitters, we observed an escape fraction of f$_{\rm esc}=14.2\pm1.9$\% at a radius of 30 kpc, twice the f$_{\rm esc}$ as the typical aperture used for photometry of Ly$\alpha$ selected galaxies. Assuming the \cite{Calzetti2000} dust law, we can estimate that attenuation of Ly$\alpha$ photons due to dust is a factor $9\pm3$ (A$_{\rm Ly\alpha} = 2.4\pm0.3$). Hence, this predicts that f$_{\rm esc}$ before dust attenuation is $100/(9\pm3)$ \%. This indicates that dust attenuation is sufficient to explain that f$_{\rm esc}$ is $\sim15$\%. This may well be the case for the majority of Ly$\alpha$ selected galaxies.
However, we note that the sample of directly detected H$\alpha$-Ly$\alpha$ emitters is obviously biased towards centrally peaked, high central f$_{\rm esc}$ Ly$\alpha$ emission, as otherwise they would not have been directly detected in Ly$\alpha$. It is therefore likely that the typical SFG has a larger Ly$\alpha$ halo with a less peaked central surface brightness. The total inferred f$_{\rm esc}$ is roughly a factor two higher than f$_{\rm esc}$ measured with 3$''$ diameter apertures. This means that, if we naively assume that this factor two is constant for all 12 SFGs used in the stack, the total escape fractions of our directly detected SFGs (see Table $\ref{tab:catalog}$) range from $4-60$\%.

\section{Conclusions}
We have undertaken a panoramic matched H$\alpha$-Ly$\alpha$ NB survey to study the Ly$\alpha$ emission from a sample of 488 H$\alpha$ selected star-forming galaxies at $z=2.23$ and its dependence on radial scale and galaxy properties. Our conclusions are:

\begin{enumerate}
\item Out of the 488 observed H$\alpha$ emitters, we detect 43 sources in the NB392 imaging. 17 of these have strong Ly$\alpha$ emission (of which 5 are X-ray AGN), and we measure the UV continuum for the other 26. We put limits on the escape fractions for these 26 and the other 445 undetected HAEs. 
\item The observed Ly$\alpha$ escape fraction for individual detected sources ranges from $2 - 30$\%, and we find that these HAEs probe a wide range of star-forming systems: with masses from $3\times10^8$ M$_{\odot}$ to 10$^{11}$ M$_{\odot}$ and dust attenuations E$(B-V)=0-0.5$. We particularly note that some massive, dusty galaxies are clearly visible in Ly$\alpha$, while they have no evidence for AGN activity from X-ray observations.
\item With matched NB observations in the $J$ and $H$ band from \cite{Sobral2013}, we are also able to detect {\sc[Oii]} and {\sc[Oiii]} emission-lines for 23 and 70 galaxies respectively. Two faint Ly$\alpha$ emitting galaxies are detected in {\sc[Oiii]}. Remarkably, these have relatively high escape fraction and have the lowest stellar mass and highest Ly$\alpha$ EW from our sample.
\item While Ly$\alpha$ morphologies of individual X-ray selected AGN are typically circularly symmetric, tracing the rest-frame UV light, the Ly$\alpha$ morphologies of star-forming galaxies are more irregular. Ly$\alpha$ can appear to be offset from the rest-frame UV and H$\alpha$ emission, and also more extended, although we detect this only significantly for AGN.
\item Both for the individually detected sources and by stacking, we confirm existing trends between the Ly$\alpha$ escape fraction and SFR and dust attenuation and explore how they are extended to a larger range of parameter space - particularly more massive and dusty galaxies. Ly$\alpha$ escape increases for galaxies with lower SFR and low dust attenuation, albeit with significant scatter. The trend between f$_{\rm esc}$ and dust attenuation resembles that observed in local galaxies. The escape fraction shows bimodal behaviour with UV slope: f$_{\rm esc}$ increases for the bluest and reddest galaxies and is at a minimum UV slopes $\beta \sim -0.5$. At the highest masses and dust attenuations, we detect individual galaxies with f$_{\rm esc}$ much higher than the typical values from stacking, indicating significant scatter in the values of f$_{\rm esc}$ at the highest masses and dust attenuations.
\item We interpret apparently contradictory reported numbers of the escape fraction by various studies using our observed trends with galaxy properties. Deeper surveys over a smaller area infer a higher escape fraction as the SFRs of their galaxies are lower in general. As these studies miss rarer galaxies with higher SFR and a higher dust attenuation, the inferred correlation between f$_{\rm esc}$ and dust content steepens. Studies based on UV selected galaxies report higher numbers of f$_{\rm esc}$ because the galaxies in their sample are biased towards galaxies with little dust, which tend to have higher f$_{\rm esc}$, except if their typical SFR is higher. 
\item The stack of our directly detected H$\alpha$-Ly$\alpha$ emitters reveals that Ly$\alpha$ is more extended than the UV and H$\alpha$, with an extent of at least $\sim30$ kpc. For Ly$\alpha$, the radius at which 90 \% of the flux is retrieved is r$_{90} = 31.3\pm1.5$ kpc, roughly 1.6 times larger than H$\alpha$. The median escape fraction for this biased sample increases up to $14.2\pm1.9$\% at least. We suggest that the missing Ly$\alpha$ photons can be fully explained by dust absorption. At the typical apertures used for slit spectroscopy (e.g. 1$''$), only half of the Ly$\alpha$ flux is recovered, even for the sample which is biased towards centrally peaked Ly$\alpha$ emission. This means that slit spectroscopy might miss a significant fraction of Ly$\alpha$.
\item By stacking our full sample of $z=2.23$ SFGs, we find that Ly$\alpha$ extends up to at least 3$''$ (with r$_{90}=36.0\pm3.8$ kpc) in radial distance at 3$\sigma$, corresponding to $\sim 20$ kpc (30 kpc, 2$\sigma$), roughly twice the scale of star formation as traced by H$\alpha$ and by the UV, with a Ly$\alpha$ surface brightness of $\sim 6\times10^{-19}$ erg s$^{-1}$ cm$^{-2}$ arcsec$^{-2}$. As a result, the Ly$\alpha$ escape fraction of our full sample of H$\alpha$ emitters continues to increase with radius, up to at least 1.6$\pm0.5$ \% at 30 kpc distance from the centre. With the current depth, it is not yet possible to constrain the maximum extent of Ly$\alpha$.
\end{enumerate}

Two important questions regarding the escape of Ly$\alpha$ remain. First, what is the total f$_{\rm esc}$ for the typical SFG at $z=2.23$? Deeper Ly$\alpha$ observations are required to fully answer this by measuring the full extent of Ly$\alpha$. Second, properties which are not studied in this survey are likely equally or more important (for example the {\sc Hi} covering fraction and geometry, e.g. \citealt{Scarlata2009}; {\sc Hi} column density, \citealt{Henry2015}; outflows, \citealt{Atek2008}; ionisation state, \citealt{Hayes2014}). From the significant scatter in the relations and from the bimodal behaviour in for example the relation between f$_{\rm esc}$ and dust content and UV slope, it is clear that multiple mechanisms are likely together responsible for determining the Ly$\alpha$ escape fraction and its relation to extended Ly$\alpha$ emission. 

It is crucial to improve the measurements of dust attenuation and to obtain precise measurements of redshifts and velocity offsets between Ly$\alpha$ and H$\alpha$, as these are responsible for the largest systematic errors which can go up to an order of magnitude when combined in unfortunate ways. However, the major downside is that this requires large, spectroscopic samples over a wide wavelength regime. Additional deeper Ly$\alpha$ NB observations are required in order to eliminate the biases caused by binning and stacking, and to be able to study the relation between galaxy properties and spatial extent of Ly$\alpha$ on a source-by-source basis.

\section*{Acknowledgments}
We thank the anonymous referee for constructive comments and suggestions which have improved the quality of this work. JM acknowledges the support of a Huygens PhD fellowship from Leiden University. DS and JM acknowledge financial support from the Netherlands Organisation for Scientific research (NWO) through a Veni fellowship, and DS from FCT through a FCT Investigator Starting Grant and Start-up Grant (IF/01154/2012/CP0189/CT0010) and from FCT grant PEst-OE/FIS/UI2751/2014. IO acknowledges support from the European Research Council (ERC) in the form of Advanced Investigator Programme, COSMICISM, 321302. HR acknowledges support from the ERC Advanced Investigator program NewClusters 321271. IRS acknowledges support from STFC (ST/L00075X/1), the ERC Advanced Investigator programme DUSTYGAL 321334 and a Royal Society/Wolfson Merit Award. APA acknowledges support from the Funda\c{c}ao para a Ci\^{e}ncia e para a Tecnologia (FCT) through the Fellowship SFRH/BD/52706/2014.

Based on observations made with the Isaac Newton Telescope (proposals 2013AN002, 2013BN008, 2014AC88, 2014AN002, 2014BN006, 2014BC118) operated on the island of La Palma by the Isaac Newton Group in the Spanish Observatorio del Roque de los Muchachos of the Instituto de Astrof\'isica de Canarias. We acknowledge the tremendous work that has been done by both COSMOS and UKIDSS UDS/SXDF teams in assembling such large, state-of-the-art multi-wavelength data-sets over such wide areas, as those have been crucial for the results presented in this paper. The sample of H$\alpha$ emitters is publicly available from \cite{Sobral2013}.

 We have benefited greatly from the public available programming language {\sc Python}, including the {\sc numpy, matplotlib, pyfits, scipy} \citep{SCIPY,MATPLOTLIB,NUMPY} and {\sc astropy} \citep{ASTROPY} packages, the imaging tools {\sc SExtractor, Swarp} and {\sc Scamp} \citep{Bertin1996,SCAMP,Bertin2010} and the {\sc Topcat} analysis program \citep{Topcat}.




\bibliographystyle{mnras}

\bibliography{bib_LAEevo.bib}




\bsp	
\label{lastpage}
\end{document}